\let\texyear\year
\documentclass{ieeeaccess}
\let\ieeeaccessyear\year
\let\year\texyear
    \usepackage{orcidlink}
\let\year\ieeeaccessyear
\definecolor{accessblue}{RGB}{0,105,154}

\usepackage{amsmath,amssymb,amsfonts}
\usepackage{algorithmic}
\usepackage{graphicx}
\usepackage{textcomp}
\usepackage{hyperref}
\usepackage{booktabs}
\usepackage{array}
\usepackage{tabularx}
\usepackage[T1]{fontenc}
\usepackage[utf8]{inputenc}

\usepackage{xcolor}
\newcommand{\new}[1]{{#1}}

\DeclareGraphicsExtensions{%
.pdf,.PDF,%
.png,.PNG,%
.jpg,.mps,.jpeg,.jbig2,.jb2,.JPG,.JPEG,.JBIG2,.JB2}

\def\BibTeX{{\rm B\kern-.05em{\sc i\kern-.025em b}\kern-.08em
             T\kern-.1667em\lower.7ex\hbox{E}\kern-.125emX}}

\begin{document}
\history{Received 6 November 2025, accepted 5 December 2025, date of publication 16 December 2025, date of current version 30 December 2025}
\doi{10.1109/ACCESS.2025.3644952}

\title{The State-of-the-Art in Lifelog Retrieval: \\A Review of Progress at the ACM Lifelog Search Challenge Workshop 2022-24}
\author{\uppercase{Allie Tran}\orcidlink{0000-0002-9597-1832}\authorrefmark{1},
	\uppercase{Werner Bailer}\orcidlink{0000-0003-2442-4900}\authorrefmark{2} ,
	\uppercase{Duc-Tien Dang-Nguyen}\orcidlink{0000-0002-2761-2213}\authorrefmark{3},
	\uppercase{Graham Healy}\orcidlink{0000-0001-6429-6339}\authorrefmark{1},
	\uppercase{Steve Hodges}\orcidlink{0000-0001-9314-7762}\authorrefmark{4},
	\uppercase{Björn \TH{}ór Jónsson}\orcidlink{0000-0003-0889-3491}\authorrefmark{5},
	\uppercase{Luca Rossetto}\orcidlink{0000-0002-5389-9465}\authorrefmark{1},
	\uppercase{Klaus Schoeffmann}\orcidlink{0000-0002-9218-1704}\authorrefmark{6},
	\uppercase{Minh-Triet Tran}\orcidlink{0000-0003-3046-3041}\authorrefmark{7},
	\uppercase{Lucia Vadicamo}\orcidlink{0000-0001-7182-7038}\authorrefmark{8},
	\uppercase{Cathal Gurrin}\orcidlink{0000-0003-2903-3968}\authorrefmark{1}
	\address[1]{Dublin City University, Ireland}
	\address[2]{Joanneum Research, Austria}
	\address[3]{University of Bergen, Norway}
	\address[4]{Lancaster University, UK}
	\address[5]{Reykjavik University, Reykjavík, Iceland}
	\address[6]{Klagenfurt University, Austria}
	\address[7]{University of Science, VNU-HCM, Vietnam}
	\address[8]{Institute of Information Science and Technologies, National Research Council (CNR), Italy}
}
\tfootnote{This publication has emanated from research conducted with the financial support of Research Ireland under grant number 13/RC/2106 P2 at the ADAPT Research Centre at Dublin City University,
	% Werner
	and has been partially funded by the European Union's Horizon Europe programme under grant agreement n$^\circ$ 101070250 XRECO, and n$^\circ$ 101092612 SUN,
	%Luca
	by the Swiss National Science Foundation via project MediaGraph, grant number 202125, 
    by the Italian Ministry of University and Research under the EU NextGenerationEU programme (PRIN 2022 PNRR P2022BW7CW, MUCES project, CUP: B53D23026090001), 
    and by Icelandic Research Fund under Grant 239772-051.
    This manuscript is made available under a Creative Commons CC-BY Licence.
}
\markboth
{A. Tran \headeretal: State-of-the-Art in Lifelog Retrieval: A Review of Progress}
{A. Tran \headeretal: State-of-the-Art in Lifelog Retrieval: A Review of Progress}

\corresp{Corresponding author: Allie Tran (e-mail: allie.tran@dcu.ie).}
\begin{abstract}
The ACM Lifelog Search Challenge (LSC) is the only long-running benchmark that evaluates lifelog retrieval systems through real-time, human-in-the-loop interaction.
This paper presents a longitudinal analysis of technical and interaction design progress at LSC from 2022 to 2024, conducted under fixed datasets, tasks, and evaluation protocols.
Unlike prior surveys of lifelogging research, our study leverages competitive, synchronous evaluations to examine how retrieval architectures, interface design choices, and interaction strategies directly influence task performance.
Through a comparative analysis of known-item search, question answering, and ad-hoc retrieval tasks, we identify a clear paradigm shift from concept-based pipelines toward embedding-driven and LLM-supported retrieval systems. 
We show how these shifts affect retrieval effectiveness, response time, and user behaviour, as well as provide empirical insight into how retrieval and interface design choices manifest in performance during interactive search. Beyond documenting emerging techniques, we expose evaluation artefacts such as performance variability across system instances and user familiarity effects, raising important considerations for future interactive benchmarks.

\end{abstract}

\begin{keywords}
	Analytics, Interactive Search, Lifelog, Multimodal Retrieval, Benchmarking
\end{keywords}

\def\thevol{13}
\def\theyear{2025}

\titlepgskip=-15pt

\maketitle
\section{Introduction}\label{sec:introduction}

Lifelogging~\cite{gurrin2014lifelogging} refers to the practice of recording aspects of daily life through wearable and other mobile devices, creating a rich and continuous record of personal experiences. As this wealth of data grows, so does the need for retrieval systems that help users to search, explore, and make sense of their past. Whether for memory enhancement~\cite{berry2007use, harvey2016remembering}, healthcare~\cite{signal2017children} or lifestyle monitoring~\cite{nguyen2016recognition,wilson2018use,tegegne2024daily}, lifelog retrieval is essential for subsequently transforming raw data into meaningful insights.

Lifelog data presents unique retrieval challenges that distinguish it from traditional multimedia search tasks such as TRECVID~\cite{2024trecvidawad} or ImageCLEF~\cite{ionescu2024overview}.
In those settings, search is asynchronous, and performance can be evaluated automatically. The first challenge is scale. Lifeloggers generate a high volume of images, video segments, and sensor readings daily, making it difficult to locate specific events~\cite{gurrin2014lifelogging}. The second challenge is personal context. Unlike conventional multimedia search, lifelog data is tied to personal memories and emotions~\cite{hodges2006sensecam,chaudhari2007privacy,dang2017building}. Retrieval systems must bridge this gap to return meaningful results. The third challenge is multi-modality~\cite{doherty2007multimodal}. As mentioned above, lifelogs combine many different types of information, requiring retrieval systems to integrate diverse sources to be effective and comprehensive.

Beyond these technical challenges, data availability remains a major obstacle. Large-scale studies require extensive long-term datasets, yet finding lifeloggers willing to record their daily lives presents major recruitment and retention challenges~\cite{gurrin2021experiments}. Consequently, only a handful of shared lifelog datasets exist~\cite{chung2022real}, forcing \new{researchers to search for events in data that are not their own}.
This lack of personal context makes it difficult to accurately interpret events. Without lived experience, searches may be substantially inaccurate or contextually irrelevant, lacking the internal cues that lifeloggers naturally rely on. For example, a photo of a meal might appear in the dataset, but without additional context, was it a significant occasion or just an ordinary lunch?

Another fundamental challenge is the nature of human memory, which is often fragmented, distorted, or shaped by emotions rather than factual details~\cite{chen2010augmenting}. This makes query formulation particularly difficult.
Combined with the noise and unstructured nature of passively collected data, this makes iterative, interactive search essential.
This manner of search also closely mirrors how people naturally search for information -- through trial and adjustment. Effective user interface (UI) and user experience (UX) design becomes crucial~\cite{dang2018lse2018}. Good interfaces allow users to refine searches and explore their lifelogs intuitively, even without precise recall or first-hand experience.

To address these challenges, the ACM Lifelog Search Challenge (LSC) was established to advance interactive lifelog retrieval through a competitive, real-time evaluation framework The ultimate goal is to move lifelog retrieval beyond basic data processing by creating tools that support reflection, true recall, and deeper engagement with personal histories. The LSC has been held annually since 2018, with the latest edition, LSC'24~\cite{lsc2024}, taking place at ACM ICMR 2024, in Phuket, Thailand.
This paper reviews the key technical advances demonstrated in the past three LSC editions (2022–2024)~\cite{lsc2022,lsc2023,lsc2024}.
We analyse trends in retrieval and interaction techniques, evaluate comparative system performance, and reflect on emerging research directions in lifelog search. Rather than proposing novel methods, this paper provides a structured analysis of the field through the lens of a recurring, shared evaluation challenge.

\new{Although this study focuses on the LSC series, its longitudinal design across identical datasets and tasks provides a uniquely controlled lens for observing algorithmic and interaction-design evolution.
Comparable long-term, human-in-the-loop benchmarks are rare in multimedia retrieval. Analysing LSC systems, therefore, offers insight into how the broader field transitions from concept-based to embedding- and LLM-driven paradigms under identical experimental conditions.}

The remainder of the paper is structured as follows: Section~\ref{sec:workshops} describes the LSC competition setup, including the dataset, tasks, and evaluation procedures. Section~\ref{sec:advances} summarises the retrieval and interaction techniques employed by the systems submitted to the past three editions of the challenge. In Section~\ref{sec:comparison}, we present a comparative analysis of system performance across tasks and years. Finally, Section~\ref{sec:conclusion} concludes with reflections on current limitations and outlines promising directions for future lifelog retrieval research.

\section{The ACM Lifelog Search Challenge Workshops}\label{sec:workshops}
LSC~\cite{lsc2022,lsc2023,lsc2024} is an annual challenge that is part of the ACM International Conference on Multimedia Retrieval (ICMR). The challenge was first introduced in 2018 and has attracted the highest participation among similar lifelogging challenges.
The focus of the LSC is to evaluate the performance of interactive lifelog search engines \textit{synchronously}. Different systems compete with each other in a live environment (on-site or remote); participants are given a set of queries and a limited time to find the relevant images.

Over the years, both the types of tasks and the size of the datasets used have evolved, making the competition increasingly challenging.

LSC'18 started with a single type of task, \textbf{known-item search (KIS)}, where participants are given a textual query describing the content they need to find. This query advances in stages to imitate gradual memory recall (see Section~\ref{sec:kis} and Table~\ref{tab:task_example}). Participants submit a single relevant image to the host server~\cite{DRES}, which manages the time limit and assesses accuracy against the ground truth. Some images from the corresponding event in the dataset, which we refer to as ground truth, are shown in Figure~\ref{fig:groundtruth-example}. In subsequent years, additional task types were introduced, specifically \textbf{Ad-hoc} and \textbf{Question Answering (QA)} tasks, which are described in Section~\ref{sec:tasks}.

\begin{table}[t!]
	\caption{An example KIS task from LSC'20~\cite{tran2023comparing}. Task 1, with its temporally advancing descriptors, was revealed at 30-second intervals. After 150 seconds, the full description is shown for another 150 seconds until the end of the task.}\label{tab:task_example}
	\begin{tabularx}{\columnwidth}{cX}
		\toprule
		\textbf{Time} & \textbf{Text}                                                                                                                                                                                                                                  \\ \midrule
		0s            & I was building a computer alone in the early morning on a Friday\ldots                                                                                                                                                                         \\\midrule
		30s           & I was building a computer alone in the early morning on a Friday at a desk\dots                                                                                                                                                                \\\midrule
		60s           & I was building a computer alone in the early morning on a Friday at a desk with a blue background\dots                                                                                                                                         \\\midrule
		90s           & I was building a computer alone in the early morning on a Friday at a desk with a blue background. Sometimes I needed to refer to the manual\dots                                                                                              \\\midrule
		120s          & I was building a computer alone in the early morning on a Friday at a desk with a blue background. Sometimes I needed to refer to the manual. I remember some Chinese posters on the desk background\dots                                      \\\midrule
		150s          & I was building a computer alone in the early morning on a Friday at a desk with a blue background. Sometimes I needed to refer to the manual. I remember some Chinese posters on the desk background. I was in Dublin City University in 2015. \\
		\bottomrule
	\end{tabularx}%
\end{table}

\subsection{Datasets}
Building upon our previous analysis of LSC progress in 2021~\cite{tran2023comparing}, the LSC introduced a new and substantially larger dataset for the 2022 and subsequent campaigns.
The dataset consists of 18 months of lifelog data (a substantial increase over the 85 days in the preceding dataset~\cite{gurrin2019test}) from January 2019 to June 2020, making it \textit{one of the longest, richest and most comprehensive multimodal lifelog datasets available to the research community}. It integrates images, biometrics, spatial and semantic data, enabling deep exploration of lifelog retrieval techniques.
Key features of the dataset include:
\begin{itemize}
	\item \textit{Core image dataset}: 725,000 first-person point-of-view lifelog images captured by a Narrative Clip device, fully redacted in 1024 $\times$ 768 resolution. Anonymisation was applied through a combination of manual and semi-automated blurring of faces and legible text. Private content that is not appropriate for public release is also removed. The images are stored in a hierarchical directory structure, with each directory representing a day and each image file named by its timestamp.
	\item \textit{Metadata}: a CSV file containing the following information for every minute: timestamp, physical activities, biometrics. For example, heart rate, steps, and sleep status are provided. Moreover, GPS coordinates are provided to track the location of the lifelogger.
	\item \textit{Visual concepts}: Two types of visual concepts are included in the metadata file: objects and scenes. The objects are detected using an object detection model trained on the Common Objects in Context (COCO) dataset, a standard resource in computer vision research, which has 80 object classes.\footnote{For a full list of objects, please refer to~\url{https://github.com/amikelive/coco-labels/blob/master/coco-labels-2014_2017.txt}.} The scenes are detected using a scene recognition model trained on the Places365 dataset~\cite{zhou2014learning}. A total of 102 scene labels are detected, such as `waiting in line', `working', and `open area'.\footnote{For a full list of scenes, please refer to~\url{https://github.com/CSAILVision/places365/blob/master/labels_sunattribute.txt}} The bounding boxes of the detected objects and confidence scores are provided for each detected concept. In addition, optical character recognition (OCR) outputs are also provided for the associated images.
	\item \textit{Additional semantic locations}: A supplementary metadata file, provided by Tran et al.~\cite{tran2023vaisl} for the LSC'23 campaign, contains semantic names for lifelogger location (e.g., `home', `Dublin City University', `Zurich Airport', etc.).
	\item \textit{Additional flight data}: flight locations as [departing airport, arrival airport] pairs are provided by the Voxento~\cite{alateeq2023voxento} developer, who was also a participant in LSC'23.
\end{itemize}

\Figure[t!][width=0.99\columnwidth]{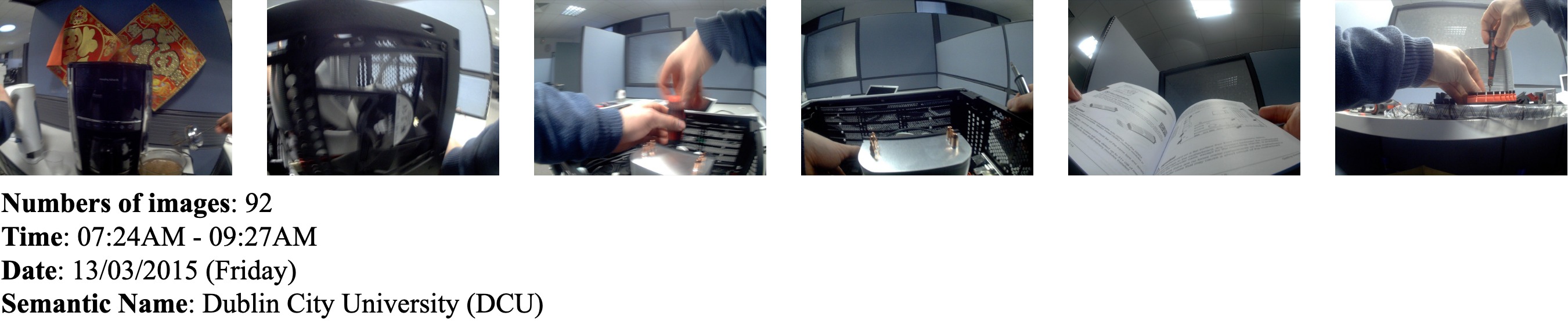}{Some lifelog images as ground truth from Task 1 in Table~\ref{tab:task_example}\label{fig:groundtruth-example}}

This dataset provides a valuable resource of information for lifelog retrieval research, enabling the development of advanced multimodal retrieval techniques. It is available upon request from the LSC organisers on the LSC website~\footnote{\url{http://lifelogsearch.org/lsc/2022/lsc_data/}}.
All identifiable faces and personal documents in the dataset were blurred, masked, or omitted using a combination of automated and manual methods.
The process of creating the dataset has been reviewed by the Dublin City University ethics board and complies with GDPR principles. Ongoing efforts aim to balance research utility with the protection of participant privacy.

\subsection{Tasks}
\label{sec:tasks}
As previously mentioned, the LSC has three types of tasks with textual queries: known-item search (KIS), ad-hoc search, and question answering (QA), see Table~\ref{tab:tasks}.
These tasks were created to reflect authentic information needs, often incorporating deliberate ambiguities to closely simulate realistic search scenarios. For tasks other than KIS, a complete ground truth is not created in advance, but human judges assess submissions in real-time.
In LSC'24, an assessment process was introduced to enhance the clarity of queries and consistency of relevance judgments. This process was adapted from the Video Browser Showdown (VBS) competition~\cite{icmr23quality}, and included a structured briefing session for query review and revision, alongside a dry-run phase for testing open-set queries. These measures aimed to minimise ambiguity and ensure consistent, synchronous relevance judgments by multiple assessors.

\begin{table*}[ht]
	\caption{Overview of the tasks in the LSC.}\label{tab:tasks}
	\centering
	\begin{tabular}{llcccc}
		\toprule
		\textbf{Task} & \textbf{Description} & \textbf{Revealed at Intervals} & \textbf{Time Limit} & \textbf{Maximum Correct Submissions} & \textbf{Form of Answer} \\ \midrule
		KIS           & Known-Item Search    & 30s                            & 300s                & 1                                    & Image                   \\
		Ad-hoc        & Ad-hoc Search        & N/A                            & 180s                & Unlimited                            & Image                   \\
		QA            & Question Answering   & N/A                            & 180s                & 1                                    & Text                    \\
		\bottomrule
	\end{tabular}
\end{table*}

\subsubsection{Known-Item Search}
\label{sec:kis}
Known-Item Search (KIS) is the oldest task in the LSC, where participants are given a textual query describing one or more scenes and are required to retrieve \textit{one} of a predefined set of relevant images within a specified time limit.
The query is formulated as a series of temporally advancing descriptors revealed at 30-second intervals (see, e.g., Table~\ref{tab:task_example}). The task simulates real-life scenarios where individuals gradually recall details over time. There is no limit on the number of incorrect submissions a participant can make before a correct submission, which determines the successful completion of the task for that participant. The time limit for this task is 300 seconds (5 minutes).

\subsubsection{Ad-hoc Search}
Ad-hoc search has been a common task in many retrieval benchmarking campaigns~\cite{lokovc2022task}, including VBS~\cite{vadicamo2024evaluating}, and the Japanese National Institute of Informatics NTCIR Lifelog Task~\cite{zhou2023overview}. However, it was only in 2022 that ad-hoc search tasks were introduced to the LSC\@.
In these tasks, participants must submit as many images as possible relevant to the given textual query within a specified time limit. For example, a query used in the latest LSC competition was: `\textit{Find examples of when I was wearing my red Vietnam football shirt}'. Because no ground truth is provided in advance, human judges are required to assess the relevance of the submissions.
There is no predefined limit on the number of submissions (correct or incorrect) a participant can make, and the time limit is 180 seconds.

\subsubsection{Question Answering}
The concept of lifelog-based question answering was initially explored through the LLQA benchmark~\cite{tran2022llqa}, which demonstrated the potential of lifelog QA for supporting memory recall and underscored its challenges in requiring reasoning across visual and temporal data.
Concurrently, the QA task was introduced in LSC'22, initially requiring participants to retrieve an image that answered a given question.
Since 2023, the task has evolved, now requiring participants to submit a text answer instead. For instance, an actual task from the LSC'24 competition was: `\textit{How many times did I visit an outdoor/farmers market in February 2020?}'. Similar to ad-hoc search, the correctness of the answers is evaluated by human judges in real-time. Additionally, as in KIS tasks, there is no limit on the number of incorrect submissions a participant can make before a correct answer is submitted. Participants are given a time limit of 180 seconds to complete the task.

\subsection{Synchronous Evaluation}\label{sec:evaluation}
The LSC uses a live scoring system, DRES~\cite{DRES}, which allows the user of each participating system to submit their queries and receive the judgment in real-time.

For KIS tasks, the evaluation metrics are based on the time taken to submit the correct image and the number of wrong submissions. The score is calculated as follows:

\begin{equation}
	\text{score}_{{KIS}} = \max\left(0,\ 100 - 50 \cdot \frac{t}{T} - 10 \cdot w\right)
	\label{eq:score}
\end{equation}
\noindent
where:
\begin{itemize}
	\item $t$ is the time taken,
	\item $T$ is the time limit,
	\item $w$ is the number of wrong submissions.
\end{itemize}
If the task is not solved, the score is 0.  Additionally, to discourage excessive submissions and to account for low precision in the submitted answers, the scoring formula includes a penalty of 10 points for each incorrect submission. This balance rewards both precision and efficiency, with successful submissions made at the five-minute time limit receiving 50\% of the maximum score. To provide transparency, we also report correctness and time-to-answer metrics independently.

For the question answering tasks, the same scoring mechanism is used, but because the answer is in the form of free-form text, correctness is determined by human judges.

For ad-hoc tasks, the evaluation metric is based on a pooled set of relevant documents for each query, the number of relevant documents retrieved, and the number of wrong submissions. Human judges assess the relevance of the submitted images in real-time during the competition, ensuring all submissions are in the pool. Pooled evaluation allows scalable annotation and aligns with TREC-style\footnote{https://trec.nist.gov/} protocols.  The score is calculated as follows:

\begin{equation}
	\text{score}_{Ad\text{-}hoc} = 100 \times \frac{\text{correct}}{\text{correct} + \text{incorrect}/2} \times \frac{\text{correct}}{\text{total}}
	\label{eq:ad-hoc}
\end{equation}
where \textit{correct} is the number of relevant images submitted, \textit{incorrect} is the number of irrelevant images submitted, and \textit{total} is the total number of relevant images aggregated over all submitting teams. The number of incorrect images is divided by 2 to reduce the penalty for wrong submissions because the number of submissions in the task is unlimited, and the query hints are sometimes fuzzy. Similarly, if no relevant images are submitted, the score is 0.

These two scoring functions are popular in the information retrieval community and are employed in other video retrieval benchmarks, such as VBS~\cite{heller2022interactive}.

\subsection{Expert and Novice Users}
LSC workshops traditionally involve two types of users: experts and novices. Expert users are those with prior experience in lifelogging, familiarity with the dataset, and, typically, direct involvement in the development of the systems they are using. Novice users, on the other hand, have no prior exposure to the dataset and or the systems tested in the challenge; they are usually recruited from the conference audience to test the usability of each system in real-time scenarios. However, from 2020 to 2022, the LSCs were held virtually due to COVID-19, and no novice users participated. They were reintroduced in LSC’23 and LSC’24.

In this paper, we focus on expert users because the performance of novice users exhibits high variability, making it difficult to draw consistent or meaningful conclusions. This variability stems from differences in search strategies and system familiarity, which is compounded by limited onboarding time, making results harder to compare across participants. While this limits conclusions about first-time usability, concentrating on expert users ensures a more controlled evaluation of retrieval performance, minimising the impact of user-related variability and noise for fair system comparison.

However, the importance of UI and UX design is not diminished, as effective interfaces remain critical for users to navigate complex lifelog data in real-world applications. While this study focuses on retrieval performance rather than usability testing, future work should incorporate novice users to assess the accessibility and user-friendliness of lifelog search systems.

\subsection{Relation to Prior Benchmarks and Research}
LSC shares lineage with prior interactive retrieval benchmarks such as TRECVID\cite{2024trecvidawad}, VBS\cite{vadicamo2024evaluating}, ImageCLEF Lifelog\cite{ImageCLEFlifelog20}, and NTCIR-Lifelog\cite{zhou2025overview}, yet differs in its focus and evaluation methodology. While TRECVID and VBS primarily benchmark large-scale video retrieval and interactive search efficiency in general-purpose datasets, lifelogging tasks at ImageCLEF and NTCIR introduced retrieval within personal or daily-life contexts. However, both ImageCLEF and NTCIR Lifelog tasks follow asynchronous evaluation protocols, where systems are assessed offline on pre-defined queries and ranked outputs. 
In contrast, the Lifelog Search Challenge (LSC) adopts a real-time, interactive format in which human users actively search through lifelog data using live systems. This unique approach complements earlier benchmarks by integrating algorithmic retrieval quality with user interaction, interface design, and time-constrained performance, enabling the study of human-in-the-loop search behaviour and system responsiveness under authentic, real-world conditions.

\section{Summary of Participating System Approaches}\label{sec:advances}
\begin{table*}[t]
	\centering
	\caption{Overview of the systems that participated in LSC 2022~\cite{lsc2022} and LSC 2023~\cite{lsc2023}.}
	\label{tab:LSC22_LSC23}
	\scriptsize
	\begin{tabular}{>{\raggedright\arraybackslash}p{2cm}>{\raggedright\arraybackslash}p{0.8cm}>{\raggedright\arraybackslash}p{3.2cm}>{\raggedright\arraybackslash}p{3.2cm}>{\raggedright\arraybackslash}p{3.2cm}>{\raggedright\arraybackslash}p{2.5cm}}
		\toprule
		\textbf{System}                             & \textbf{Year} & \textbf{UI/UX Enhancements}              & \textbf{Search and Retrieval Techniques}       & \textbf{Advanced Features and Innovations} & \textbf{Embedding Models}                                  \\
		\midrule
		FIRST 3.0~\cite{hoang2022flexible}          & 2022          & Updated UI for better usability          & Query-by-example, text-image retrieval         & External search engine support             & -                                                          \\
		\midrule
		LifeSeeker 4.0~\cite{nguyen2022lifeseeker}  & 2022          & Novice-optimized interface               & CLIP-based retrieval, emotion-based indexing   & Event clustering for better search results & CLIP                                                       \\
		\midrule
		LIFEXPLORE~\cite{schoeffmann2022lifexplore} & 2022          & Improved UI, better search visualization & Semantic geo-location, temporal filtering      & Context-aware search refinement            & -                                                          \\
		\midrule
		Memento 2.0~\cite{alam2022memento}          & 2022          & UI improvements for efficiency           & Weighted CLIP ensemble retrieval               & Optimized performance over LSC'21          & CLIP ResNet50x64 and ViT-L/14                              \\
		\midrule
		MEMORIA~\cite{ribiero2022memoria}           & 2022          & Filter-rich UI, improved annotation      & Keyword-based retrieval                        & CNN-based visual annotation                & -                                                          \\
		\midrule
		MyScéal~\cite{tran2023mysceal}              & 2022          & Novice-friendly UI, better accessibility & CLIP-based text-to-image retrieval             & Improved version of top LSC'20/21 system   & CLIP ViT-L/14                                              \\
		\midrule
		vitrivr~\cite{heller2022vitrivr}            & 2022          & Dynamic sequencing, faceted filtering    & Text embedding, Boolean retrieval              & Lifelog-specific retrieval enhancements    & CLIP and W2VV++\cite{li2019w2vv++}                         \\
		\midrule
		vitrivr-VR~\cite{spiess2022multimodal}      & 2022          & VR-based search with three modes         & Built on vitrivr, visual search                & Immersive lifelog navigation               & -                                                          \\
		\midrule
		Voxento 3.0~\cite{alateeq2022voxento}       & 2022          & Voice search, UI refinements             & Speech-to-query, metadata filtering            & Hands-free search with expanded controls   & CLIP                                                       \\
		\midrule
		ELifeSeeker~\cite{nguyen2023lifeseeker}     & 2023          & Optimized UI for Q\&A tasks              & Differential networks, embedding models        & Q\&A search with knowledge-based inference & Switchable between CoCa, OpenCLIP Vit-H/14, BLIP and ALIGN \\
		\midrule
		\midrule
		FIRST~\cite{hoang2023lifelog}               & 2023          & UI optimized for generative search       & Generative query expansion, text-image search  & AI-assisted retrieval and query assistance & CLIP                                                       \\
		\midrule
		Hybrid vitrivr~\cite{spiess2023best}        & 2023          & Hybrid desktop-VR interface              & vitrivr-based multimodal retrieval             & Desktop-VR search integration              & OpenCLIP ViT-H/14                                          \\
		\midrule
		LifeGraph 3~\cite{rossetto2023multi}        & 2023          & Interactive knowledge graph UI           & Temporal, spatial, and visual clustering       & Multi-modal knowledge-based exploration    & OpenCLIP ViT-H/14                                          \\
		\midrule
		LifeInsight~\cite{nguyen2023lifeinsight}    & 2023          & UI with spatial Q\&A tools               & Embedding-based retrieval, event analysis      & Visualization support for Q\&A queries     & BLIP                                                       \\
		\midrule
		LifeLens~\cite{hordvik2023lifelens}         & 2023          & Minimalist, user-focused UI              & Built on E-LifeSeeker engine                   & Prioritizes simplicity and user efficiency & -                                                          \\
		\midrule
		LIFEXPLORE~\cite{schoeffmann2023lifexplore} & 2023          & Temporal filtering, updated UI           & Hybrid retrieval (embedding + CNN)             & Improved spatial-temporal search           & CLIP                                                       \\
		\midrule
		Memento~\cite{alam2023memento}              & 2023          & UI refinements for search control        & Cluster-based search, user-selectable models   & Improved efficiency and result diversity   & Weighted ensembles of different CLIP models                \\
		\midrule
		MEMORIA~\cite{ribeiro2023memoria}           & 2023          & Enhanced UI, detailed annotations        & Free-text graph database, event segmentation   & Structured event-based retrieval           & OpenCLIP ViT-H/14                                          \\
		\midrule
		MemoriEase~\cite{tran2023memoriease}        & 2023          & Conversational UI, user-friendly design  & Concept and embedding-based retrieval          & Interactive query feedback loop            & BLIP                                                       \\
		\midrule
		MyEachtra~\cite{tran2023myeachtra}          & 2023          & Revised UI, better query handling        & Event-centric retrieval, text-image embeddings & Enhanced version of MyScéal                & OpenCLIP ViT-H/14                                          \\
		\midrule
		Voxento 4.0~\cite{alateeq2023voxento}       & 2023          & Voice-enabled UI, improved controls      & Text-image embeddings, metadata filtering      & More efficient voice-based search          & CLIP ViT-L/14                                              \\
		\bottomrule
	\end{tabular}
\end{table*}
\begin{table*}
	\caption{Overview of the systems that participated in LSC 2024~\cite{lsc2024}.}\label{tab:systems}
	\centering
	\scriptsize
	\begin{tabular}{>{\raggedright\arraybackslash}p{2cm}>{\raggedright\arraybackslash}p{2.6cm}>{\raggedright\arraybackslash}p{2.6cm}>{\raggedright\arraybackslash}p{2.6cm}>{\raggedright\arraybackslash}p{2.6cm}>{\raggedright\arraybackslash}p{2.6cm}}
		\toprule
		\textbf{System}                              & \textbf{UI/UX Enhancements}          & \textbf{Search and Retrieval Techniques}                          & \textbf{Advanced Features and Innovations}                                               & \textbf{Embedding Models}                                                         & \textbf{LLMs}                                          \\
		\midrule
		CollaXRSearch~\cite{ly2024collaxrsearch}     & Shared virtual workspace             & Collaborative VR environment for lifelog retrieval                & Hybrid device setup, head-gaze-driven interaction                                        & BLIP and CLIP models for text-image retrieval                                     & -                                                      \\
		\midrule
		EAGLE~\cite{nguyenho2024eagle}               & Eye-tracking based UI                & Eye movement-based elimination method                             & Eye-tracking integration                                                                 & CLIP to generate initial results                                                  & -                                                      \\
		\midrule
		Exquisitor~\cite{khan2024exquisitor}         & Streamlined UI for search modes      & Conversational search, URF model                                  & Stateful, multi-turn conversational search                                               & CLIP models for conversational and relevance feedback search                      & -                                                      \\
		\midrule
		Libro~\cite{hezel2024libro}                  & Dynamic grid layout for UI           & Keyframe and shot detection                                       & Graph-based visualization and exploration                                                & EVA CLIP Vit-E~\cite{fang2023eva}                                                 & -                                                      \\
		\midrule
		LifeGraph 4~\cite{rossetto2024lifegraph}     & -                                    & Event-based clustering, multimodal knowledge graphs               & Relevance feedback and result diversification                                            & Vision-Language Models (VLMs) such as BLIP-2 and LLaVA for information extraction & -                                                      \\
		\midrule
		LifeInsight 2.0~\cite{vuong2024lifeinsight}  & Redesigned interface, side tab bar   & Temporal queries, ensemble models for image-text retrieval        & Combines CLIP and BLIP2 models                                                           & CLIP and BLIP2 models for image-text retrieval                                    & -                                                      \\
		\midrule
		LifeLens 2.0~\cite{tysse2024lifelens}        & Integrated search bar, drag-and-drop & -                                                                 & Refined UI components                                                                    & User-selectable CLIP models                                                       & -                                                      \\
		\midrule
		LifeSeeker 6.0~\cite{le2024lifeseeker}       & Enhanced user interface              & Contrastive learning for query matching                           & Support for question-answering tasks                                                     & Contrastive learning models using embeddings (IaT and TaT)                        & -                                                      \\
		\midrule
		LIFEXPLORE~\cite{rader2024lifexplore}        & Improved GUI, new query-building     & Temporal and combined query capabilities                          & Integration of OpenCLIP models                                                           & OpenCLIP models                                                                   & -                                                      \\
		\midrule
		Memento 4.0~\cite{alam2024memento}           & Chat-based QA format                 & Hierarchical event segmentation, QA pipeline                      & Use of LLMs for enhanced retrieval                                                       & Assembles of CLIP models                                                          & GPT-3.5 Turbo and Mistral7B for RAG                    \\
		\midrule
		MEMORIA~\cite{gago2024memoria}               & Intuitive interface, new components  & Event classification, image upload search                         & Integration of LLMs for querying and event description                                   & -                                                                                 & Integration of LLMs for querying and event description \\
		\midrule
		MemoriEase~\cite{tran2024memoriease}         & Conversational interface             & Conversational search, visual similarity search                   & Retrieval-Augmented Generation (RAG) with LLMs                                           & BLIP-2 model for embedding and InstructBLIP for description generation            & GPT-3.5 for RAG                                        \\
		\midrule
		MyEachtraX~\cite{tran2024myeachtrax}         & Mobile-friendly UI                   & Dual retrieval approaches                                         & Enhanced Event Reader component                                                          & OpenCLIP for text-image retrieval                                                 & LLMs and MLLMs for query parsing and post-processing   \\
		\midrule
		PraK Tool V2~\cite{vopalkova2024searching}   & Centralized architecture             & Temporally distant activity search, metadata filtering            & Scalable system architecture                                                             & -                                                                                 & -                                                      \\
		\midrule
		Retrospect~\cite{steffensen2024t@retrospect} & Streamlined UI, progress bar         & Task-specific features for LSC'24 tasks                           & Iterative user testing and feedback integration                                          & User-selectable CLIP models                                                       & -                                                      \\
		\midrule
		SnapSeek~\cite{hole2024snapseek}             & Location-clustered timeline display  & Visual similarity search                                          & Uses multiple models for embedding computation                                           & OpenCLIP, BLIP2, and BEiT-3 models for image embedding                            & -                                                      \\
		\midrule
		VitaChronicle~\cite{pagani2024vitachronicle} & User-centred design, simplified UI   & -                                                                 & Focus on UX/UI principles                                                                & -                                                                                 & -                                                      \\
		\midrule
		vitrivr~\cite{sauter2024general}             & -                                    & Boolean retrieval, full-text search                               & General-purpose retrieval stack with advanced tools                                      & OpenCLIP ViT-B/32\_xlm\_roberta\_base                                             & -                                                      \\
		\midrule
		vitrivr-VR~\cite{spiess2024spatiotemporal}   & Map-based VR interface               & Spatiotemporal queries, calendar-like temporal queries            & VR interaction for lifelog analytics                                                     & OpenCLIP for visual-text co-embedding                                             & -                                                      \\
		\midrule
		VISIONE~\cite{amato2024visione}              & GUI modifications for lifelog data   & Temporal, text-to-image, similarity, object/colour-based searches & Adaptation from video retrieval system, late fusion of three multimodal embedding models & OpenCLIP Vit-L/14 trained on LAION, CLIP2Video, and ALADIN                        & -                                                      \\
		\midrule
		Voxento-Pro~\cite{alateeq2024voxentopro}     & Chat interface for UI                & Conversational search, semantic search                            & Integration of OpenAI's Assistant and Whisper APIs                                       & OpenAI CLIP and OpenCLIP for semantic search                                      & OpenAI Assistant API and Whisper API                   \\
		\bottomrule
	\end{tabular}
\end{table*}

Lifelog retrieval systems must interpret user inputs across multiple modalities, such as free-text queries, visual examples, and contextual filters (e.g., time, location, or activity). The strategies for interpreting these inputs vary by system. \new{Because integration details are team-specific and often unpublished, our discussion focuses on observed usage patterns and system-level effects rather than implementation comparisons}.

The strategies used vary significantly across systems, ranging from rule-based filtering to embedding-based semantic search and LLM-driven query interpretation.
Tran et al.~\cite{tran2024lifelogging} present a general pipeline for LSC systems, illustrating how user inputs are processed, matched against indexed data, and refined through user interaction. This model reflects the common architecture seen in many LSC submissions and helps contextualise the multimodal input-handling approaches used in the challenge. 

The early years of LSC, from 2018 to 2021, saw a gradual progress in lifelog retrieval methods. Developments from this period are detailed in~\cite{gurrin2019invited} and~\cite{tran2023comparing}, which documented the advancements made in the initial and later years of this phase, respectively.

Since 2022, LSC has seen a steady increase in participating teams, many of whom have introduced new retrieval techniques and system designs each year. Some approaches, such as metadata filtering, temporal queries, and visual similarity search, have now become standard in the community and continue to be refined. Meanwhile, emerging trends—including embedding-based retrieval (with embeddings extracted via vision transformers or language-image transformers), large language models (LLMs), and multimodal query interfaces—are transforming lifelog search.

This section reviews the major technical advances made in LSC’22, LSC’23, and LSC’24 and provides a structured analysis. Tables~\ref{tab:LSC22_LSC23} and~\ref{tab:systems} summarise the systems across these editions, highlighting key improvements in UI and UX, search and retrieval techniques, and advanced features.

\subsection{Evolution of Lifelog Retrieval Systems}\label{sec:evolution}
\vspace{1ex}
\subsubsection{LSC'22: The Rise of CLIP-Based Embedding Models}
LSC’22 marked a significant shift toward embedding-based retrieval, with many teams integrating contrastive language–image pretraining (CLIP) models~\cite{radford2021learning} for text-to-image search, which are language-image transformers that are trained in a contrastive way to maximise text-image similarity. Systems such as MyScéal~\cite{tran2023mysceal} and LifeSeeker 4.0~\cite{nguyen2022lifeseeker} adopted CLIP to improve semantic understanding, enabling more abstract and flexible queries compared to traditional keyword-based retrieval. Memento 2.0~\cite{alam2022memento} introduced a weighted ensemble approach for CLIP integration, optimising retrieval efficiency.

User experience improvements were also a focus. MyScéal introduced a more accessible search interface, while LifeSeeker 4.0 used event clustering to enhance navigation.\ vitrivr-VR~\cite{spiess2022multimodal} introduced expanded their Virtual Reality (VR) search environment, offering three immersive interfaces for lifelog exploration. LIFEXPLORE~\cite{schoeffmann2022lifexplore} improved semantic geo-location enrichment and search result ordering to enhance retrieval accuracy. Alternative interaction methods gained traction. Voxento 3.0~\cite{alateeq2022voxento} implemented voice-based search with speech-to-query mechanisms, while FIRST 3.0~\cite{hoang2022flexible} and vitrivr~\cite{heller2022vitrivr} enhanced their retrieval pipelines by incorporating external search engines for unfamiliar concept expansion.

Overall, LSC’22 solidified embedding-based retrieval, multimodal search, and several UI/UX innovations, setting the stage for further advancements in lifelog retrieval.

\subsubsection{LSC'23: Expansion of Multimodal and Temporal Retrieval}
Building on the advances of LSC’22, the 2023 edition saw greater adoption of text-image embedding models in nearly all systems. A significant trend was the integration of temporal and spatial filtering, with systems such as LIFEXPLORE \cite{schoeffmann2023lifexplore} and Memento \cite{alam2023memento} refining event-based retrieval by incorporating context-aware search mechanisms.

Several systems underwent major enhancements:

\begin{itemize}
	\item MyEachtra~\cite{tran2023myeachtra}, an enhanced version of MyScéal, introduced event-centric retrieval to group lifelog data more effectively.
	\item MEMORIA~\cite{ribeiro2023memoria} implemented a free-text graph database with more detailed event segmentation, improving retrieval accuracy.
	\item LifeGraph 3~\cite{rossetto2023multi} pioneered knowledge graph-based multimodal exploration, allowing users to interact with hierarchical temporal, spatial, and visual clusters.
\end{itemize}

New systems also emerged with innovative retrieval paradigms. For example, Spiess et al.~\cite{spiess2023best} developed a desktop-VR hybrid system, combining web-based queries with an immersive VR-based browsing experience. Memori\-Ease~\cite{tran2023memoriease} merged concept-based and embedding-based retrieval, while LifeInsight~\cite{nguyen2023lifeinsight} introduced a spatial insight mechanism to support question-answering queries.

LSC’23 also marked the introduction of generative AI models in search engines, which became the main focus of the next edition. For instance, FIRST~\cite{hoang2023lifelog} incorporated generative query expansion, enabling users to retrieve lifelog data based on predicted concepts.

\subsubsection{LSC'24: Integration of LLMs, Conversational Search, and Advanced UI/UX}
The 2024 edition introduced LLMs and multimodal models for conversational search, enhancing query interpretation and interactive retrieval. Systems such as MemoriEase~\cite{tran2024memoriease}, Voxento-Pro~\cite{alateeq2024voxentopro}, and Memento 4.0~\cite{alam2024memento} leveraged GPT-3.5 Turbo\footnote{A variant of GPT-3.5 with enhanced performance, as documented by OpenAI on \url{https://platform.openai.com/docs/models}} and Mistral7B~\cite{jiang2023mistral}, enabling dialogue-based interactions and retrieval-augmented generation (RAG) techniques~\cite{lewis2020retrieval}. More details can be found in later sections.

Other notable advances in LSC’24 included:
\begin{itemize}
	\item Eye-tracking for query optimization: EAGLE~\cite{nguyenho2024eagle} integrated eye-tracking technology to filter irrelevant results automatically.
	\item Collaborative search in VR/AR: CollaXRSearch~\cite{ly2024collaxrsearch} introduced a shared virtual workspace, allowing multiple users to explore lifelog data together.
	\item Conversational feedback and query refinement: Systems like Exquisitor~\cite{khan2024exquisitor} implemented stateful, multi-turn conversational search, refining retrieval based on iterative user input.
\end{itemize}

Furthermore, UI and UX improvements played a major role in LSC’24, with LifeLens 2.0~\cite{tysse2024lifelens}, VitaChronicle~\cite{pagani2024vitachronicle}, and Retrospect~\cite{steffensen2024t@retrospect} focusing on user-centric design and streamlined interfaces. These systems aimed to enhance user engagement and search efficiency through intuitive navigation and interactive features.

Table~\ref{tab:systems} details the systems introduced at LSC’24.

\subsection{Key Technical Advances in Lifelog Retrieval}
We now discuss the major techniques employed in more detail.

\subsubsection{Embedding-based Retrieval}
Embedding-based retrieval approaches have transformed lifelog search by mapping both text and images into a shared high-dimensional semantic space. Cosine similarity between the search query and the images is directly used to rank the results. Unlike traditional concept-based retrieval, which relies on explicit labels (e.g., `beach' or `car'), embeddings capture deeper contextual meanings, enabling the retrieval of relevant data even when exact keywords are missing. For instance, users can issue abstract queries like `a relaxing afternoon with friends' and receive semantically related images and videos.

This technique was mentioned in the previous analysis of LSC'21~\cite{tran2023comparing}, and it has become the dominant approach since LSC'22. Text and image embeddings are trained using large datasets of paired image-text data, allowing models such as CLIP~\cite{radford2021learning}, BLIP~\cite{li2022blip}, and BLIP-2~\cite{li2023blip} to understand complex relationships. CLIP aligns text and image embeddings to maximise their agreement for correct pairs, supporting zero-shot queries. BLIP and BLIP-2 build upon this by improving the visual question-answering and captioning capabilities, making them ideal for multimodal queries. Some optimisation methods, such as KNN search~\cite{emysceal2022} and FAISS~\cite{johnson2019billion}, can be used to speed up the search as in the case of LIFEXPLORE~\cite{schoeffmann2023lifexplore} and Memento~\cite{alam2023memento}. Although there is some effort in fine-tuning the embedding models~\cite{tran2022exploration}, large-scale pre-trained models are generally more robust and are often used, either directly or in a weighted ensemble~\cite{alam2023memento}.

At LSC'24, several systems harnessed these embedding models to enhance retrieval performance. LifeInsight 2.0~\cite{vuong2024lifeinsight} combined CLIP and BLIP-2 in an ensemble configuration to perform robust image-text retrieval, particularly in temporal queries.
SnapSeek~\cite{hole2024snapseek} employed OpenCLIP~\cite{ilharco_gabriel_2021_5143773}, BLIP-2, and BEiT-3~\cite{wang2022image} to handle queries related to visually similar events. VISIONE~\cite{amato2024visione} used a late fusion approach to combine the results obtained with three embedding models, namely OpenCLIP ViT-L/14~\cite{radford2021learning}, CLIP2Video~\cite{fang2021clip2video}, and  ALADIN~\cite{messina2022aladin}.  CollaXRSearch~\cite{ly2024collaxrsearch} enhanced collaborative VR retrieval by leveraging text-image embeddings to support shared exploration within a virtual workspace.
Exquisitor~\cite{khan2024exquisitor} integrated conversational feedback using CLIP to refine searches iteratively through user input.

While many systems relied on retrieving images individually, several recent approaches extended these embeddings to incorporate temporal context by operating over structured segments of lifelog data. Rather than treating each image in isolation, systems like MyEachtra~\cite{tran2023myeachtra}, Memento 4.0~\cite{alam2024memento}, and MemoriEase 2.0~\cite{tran2024memoriease} performed hierarchical or content-based segmentation to construct event-level representations, where visual features from temporally adjacent frames were aggregated. These event embeddings served as the atomic units for retrieval or reasoning, enabling systems to interpret queries involving longer activities or complex temporal dependencies. This shift toward event-centric representations reflects a growing interest in modelling lifelog data in ways that align more closely with human memory and real-world activity flow. Such a trend extends beyond lifelogging and is also evident in the wider multimedia retrieval community, as discussed in~\cite{schall2024interactive}.

Embedding-based methods are particularly useful for ad-hoc and open-ended queries, where precise keywords are unavailable. We expect future systems to extend this approach by incorporating additional context, such as biometric data, to enhance retrieval performance further.

\subsubsection{Large Language Models, Conversational Search, and Question Answering}
The adoption of LLMs in LSC’24 marked a major milestone in lifelog retrieval. LLMs have introduced powerful language understanding capabilities to lifelog retrieval systems. By pretraining on vast corpora of text, LLMs can interpret complex, natural language queries and support dialogue-based search. In lifelog search, this means users can refine their searches through follow-up questions or clarifications, making the retrieval process more interactive and personalised.

Retrieval-augmented generation (RAG)~\cite{lewis2020retrieval} is commonly used to improve LLM performance by grounding responses in relevant lifelog entries retrieved in real-time. This prevents the LLM from generating answers based solely on pre-existing knowledge, instead forcing it to reference actual lifelog data. QA pipelines in some systems further segment lifelog data into events, providing structured summaries and direct answers.

Examples of LLMs used in LSC'24 systems include GPT-3.5 Turbo\footnote{A variant of GPT-3.5 with enhanced performance, as documented by OpenAI on \url{https://platform.openai.com/docs/models}} and Mistral7B~\cite{jiang2023mistral}. These models enable conversational search interfaces, as seen in MemoriEase~\cite{tran2023memoriease}, Exquisitor~\cite{khan2024exquisitor}, and Voxento-Pro~\cite{alateeq2024voxentopro}. Memento 4.0~\cite{alam2024memento} leveraged GPT-3.5 Turbo and Mistral7B in its QA pipeline, providing structured answers based on event segmentation. Meanwhile, Voxento-Pro~\cite{alateeq2024voxentopro} implemented a voice-enabled conversational interface using OpenAI Assistant and Whisper APIs, allowing users to issue queries through spoken natural language.

Multimodal LLMs, such as Vision-Language Models (VLMs), have also been explored for information extraction and retrieval in MyEachtraX~\cite{tran2024myeachtrax} and LifeGraph 4~\cite{rossetto2024lifegraph}. These models combine visual and textual information to provide more comprehensive lifelog search capabilities.

\subsubsection{User Interface and User Experience}
Although this study focuses on expert users for system evaluation, UI/UX innovations remain critical for ensuring that lifelog retrieval systems are accessible to a broader audience. A well-designed interface reduces cognitive load and makes it easy for users to issue complex queries with minimal effort. Features such as drag-and-drop, sidebars for filtering, and visual previews of search results have become standard. UI/UX enhancements have been a key focus in recent LSC editions:

\begin{itemize}
	\item In LSC'22, systems like MyScéal~\cite{tran2023mysceal} and vitrivr~\cite{heller2022vitrivr} introduced novice-friendly interfaces, simplifying lifelog exploration for new users.\ vitrivr-VR~\cite{spiess2022multimodal} took this further by offering a VR-based interface for immersive data exploration.
	\item LSC'23 saw the introduction of dynamic result sequencing and faceted filtering in LIFEXPLORE~\cite{schoeffmann2023lifexplore}, enhancing search result organisation.
	      MEMORIA~\cite{ribeiro2023memoria} and Memento~\cite{alam2023memento} focused on detailed image annotations and user-selectable embedding models, respectively.
	\item LSC'24 showcased a range of UI/UX improvements, with several teams participating in the challenge with the focus of UI/UX enhancements for different levels of user expertise, such as Retrospect~\cite{steffensen2024t@retrospect}, LifeLens\cite{hordvik2023lifelens}, and VitaChronicle~\cite{pagani2024vitachronicle}. MyEachtraX~\cite{tran2024myeachtrax} brought a mobile-friendly design to lifelog search to enhance accessibility. Conversational interfaces, as mentioned earlier, are gaining in popularity, allowing users to interact with lifelog systems through natural language queries. These interfaces further enhance UX by guiding users through the search process step-by-step, allowing for query refinement through dialogue rather than manual re-entry.
\end{itemize}

Additionally, immersive interfaces gained further traction, with Virtual Reality (VR) and Augmented Reality (AR) lifelog retrieval systems offering novel ways to explore lifelog data. VR allows users to `step into' their lifelogs, navigating time and space through virtual reconstructions. AR overlays lifelog information onto real-world surroundings, enabling contextual re-experiencing of past events.  VR systems typically use spatial metaphors such as timelines and maps to organise lifelog entries. Head-gaze tracking and hand gestures often replace traditional input methods, making the interaction more natural. Collaborative VR environments take this further by allowing multiple users to explore lifelog data together. Some notable VR/AR systems showcased at LSC'24 include vitrivr-VR~\cite{spiess2024spatiotemporal}, which provided a map-based VR interface for lifelog exploration, and CollaXRSearch~\cite{ly2024collaxrsearch}, which introduced a shared VR workspace for collaborative lifelog analysis.  Despite the potential of VR/AR interfaces, they face adoption challenges due to hardware requirements and user comfort. However, as VR/AR technologies mature, they may redefine how users interact with lifelog archives.

\subsubsection{New Search and Retrieval Techniques}
Several novel techniques showcased at LSC'24 extended the capabilities of lifelog retrieval systems:

\begin{itemize}
	\item \textbf{Collaborative Search:} Collaborative search enables multiple users to interact with a lifelog space simultaneously. CollaXRSearch~\cite{ly2024collaxrsearch} supported shared exploration, making it ideal for reviewing events like family vacations or collaborative data analysis tasks.
	\item \textbf{Eye-Tracking Integration:} Eye-tracking provides implicit feedback by analysing user gaze patterns during searches. For example, EAGLE~\cite{nguyenho2024eagle} used this to eliminate irrelevant results automatically, enhancing search precision without requiring manual input.
\end{itemize}
These techniques reflect the shift towards more interactive, adaptive, and user-friendly lifelog retrieval experiences.

\section{Comparative Analysis of System Performance through the Years}\label{sec:comparison}
Since the LSC has been running for several years, it is possible to compare system performance in LSC'24 with previous editions, LSC'22 and LSC'23. \new {The same dataset and tasks were used across all editions, except in 2022, when the QA task required an image as the answer rather than text.}
Some systems participated across multiple years and experimented with different approaches, giving us insights into the effectiveness of various strategies. This section focuses on trends in system performance across three key metrics: correctness of submissions, time to the first correct submission, and overall scores.
\new{We follow a top–down analytical sequence: first, per-metric trends (correctness, timing, scoring); second, query difficulty effects that help explain those trends; and finally, a focused analysis of LSC'24 to illustrate how these factors play out in the most recent edition.}

While the tasks are designed with comparable formats, we acknowledge that individual task difficulty may vary across years. To mitigate this, our comparison focuses on average scores per task type, which reduces the impact of particularly hard or easy queries on overall performance trends.

Details of the tasks for each year are provided on the LSC main website.\footnote{\url{http://lifelogsearch.org/lsc/}} Specifically, the tasks for LSC'24 are publicly available at the challenge's website.\footnote{\url{http://lifelogsearch.org/lsc/2024/resources/LSC24-Topics-Release.pdf}}

\subsection{Performance of Systems}\label{sec:performance-comparison}
The evaluation metrics used in the LSC are based on the time taken to submit the correct image and the number of correct and incorrect submissions. The number of participating teams increased from 9 in 2022 to 14 in 2023, reaching 35 in 2024. In 2024, some systems had multiple instances (e.g., SYSTEM1, SYSTEM2), which were treated as independent participants. This increased the total number of submissions and led to higher performance variability.
\new{We begin with correctness, then examine timing, and finally relate both to aggregate scores across recurring teams.}

\subsubsection{Correctness of Submissions}
Figure~\ref{fig:correct_and_incorrect_by_task_type_and_year} shows the number of correct and incorrect submissions for KIS, QA, and Ad-hoc tasks. The values are computed from a table where each row corresponds to a unique combination of team, task, and year. In KIS and QA tasks, where submissions are limited to one correct answer, we report mean values per team-task entry, aggregated across all such entries within each year. In Ad-hoc tasks, where submissions are unlimited, we show the full distribution of correct and incorrect submissions per team-task entry, also grouped by year.

For KIS tasks, the mean number of correct submissions increased from 0.78 in 2022 to 0.87 in 2023, indicating improved system strategies. However, it dropped to 0.68 in 2024, while precision decreased to 0.64, likely due to exploratory submissions from system instances with different users. Although some instances were effective, others likely introduced noise, contributing to the lower precision. QA tasks showed consistent correct submissions (0.78 in both 2023 and 2024), but precision improved slightly from 0.50 to 0.52. Ad-hoc tasks exhibited the widest variability, with some teams performing very well and others submitting many incorrect attempts. While the median correct submissions remained stable, the spread widened in 2024, suggesting that not all instances of the same core system performed consistently.

The overall precision for KIS and QA tasks was higher than for ad-hoc tasks, which consistently recorded lower precision due to their open-ended nature. The introduction of multiple system instances in 2024 increased the diversity of approaches, but also amplified the volume of incorrect submissions, impacting precision across tasks.
\new{Taken together, these correctness and precision patterns suggest that differences in \emph{timing} may explain the variability, as slower query reformulation and exploratory submissions likely contribute to it.}

\Figure[t!][width=0.99\columnwidth]{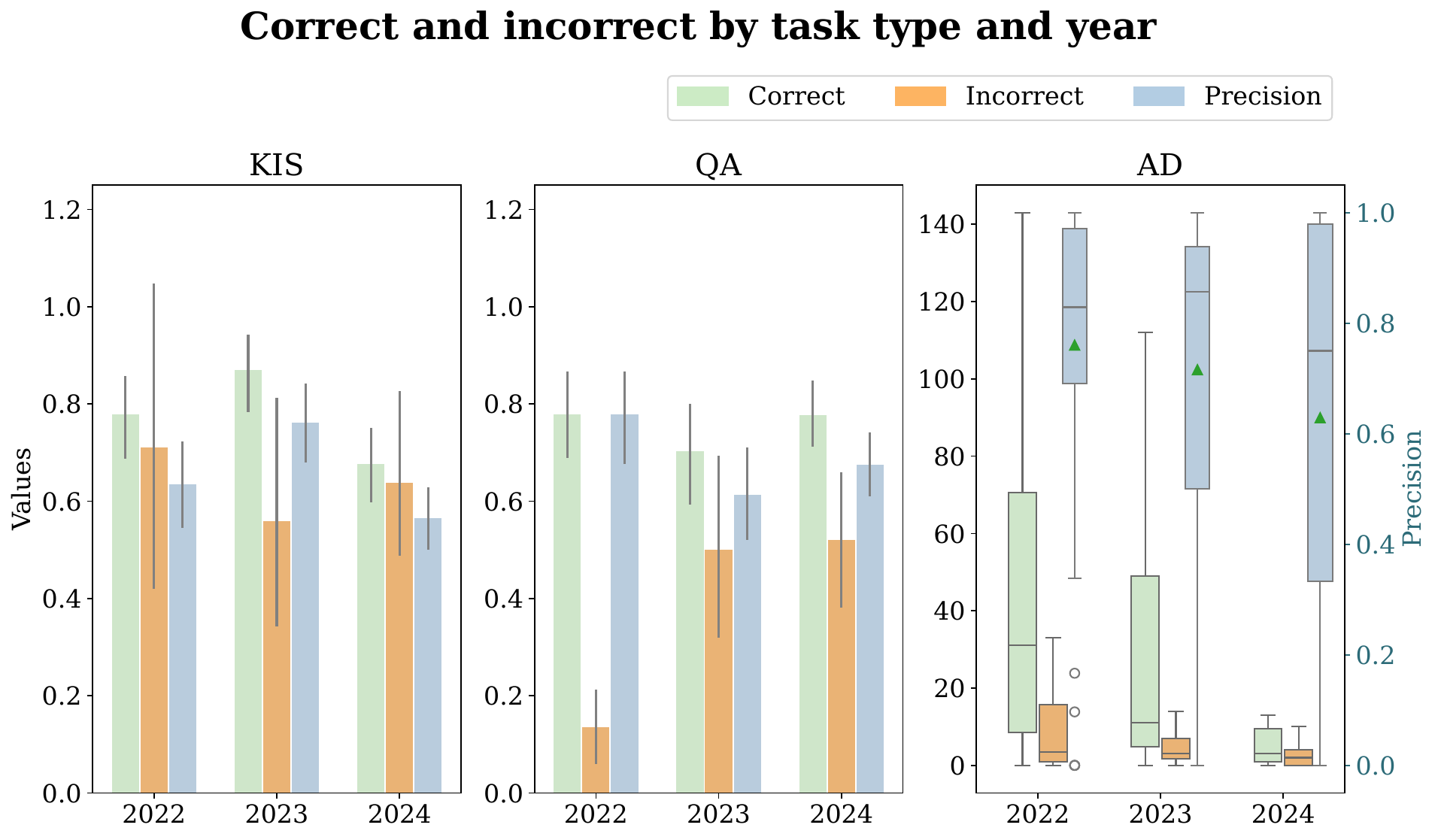}{Number of correct and incorrect submissions in LSC'22, '23, and '24.\label{fig:correct_and_incorrect_by_task_type_and_year}}

\subsubsection{Time to First Correct Submission}
\new{To contextualise the above correctness patterns, we now analyse time-to-first-correct as a proxy for query formulation and interaction efficiency.}
Figure~\ref{fig:first_correct_submission_by_all_teams_by_task_type_and_year} shows the distribution of the time taken by teams to make their first correct submission. For KIS and QA tasks, the median times remained relatively stable across years, but the spread increased in 2024 due to system instance variability. This indicates that while some instances responded quickly, others took considerably longer. Outliers in 2024 represent instances that struggled to deliver timely, correct submissions, demonstrating the user variability across instances.

For QA tasks, the spread of times was consistently large across the years, but a slight improvement in the median time in 2024 suggests some gains in system efficiency. The ad-hoc retrieval tasks showed the highest variability in 2024, with some teams submitting correct answers quickly and others taking substantially longer. This wide range of performance reflects the complexity of open-ended queries and the exploratory nature of multiple instances, with different users demonstrating varying levels of success.
\new{These timing effects help explain the larger precision differences observed above and prepare for our comparison of \emph{scores}, which combine correctness and timing in the official evaluation}

\Figure[t!][width=0.99\columnwidth]{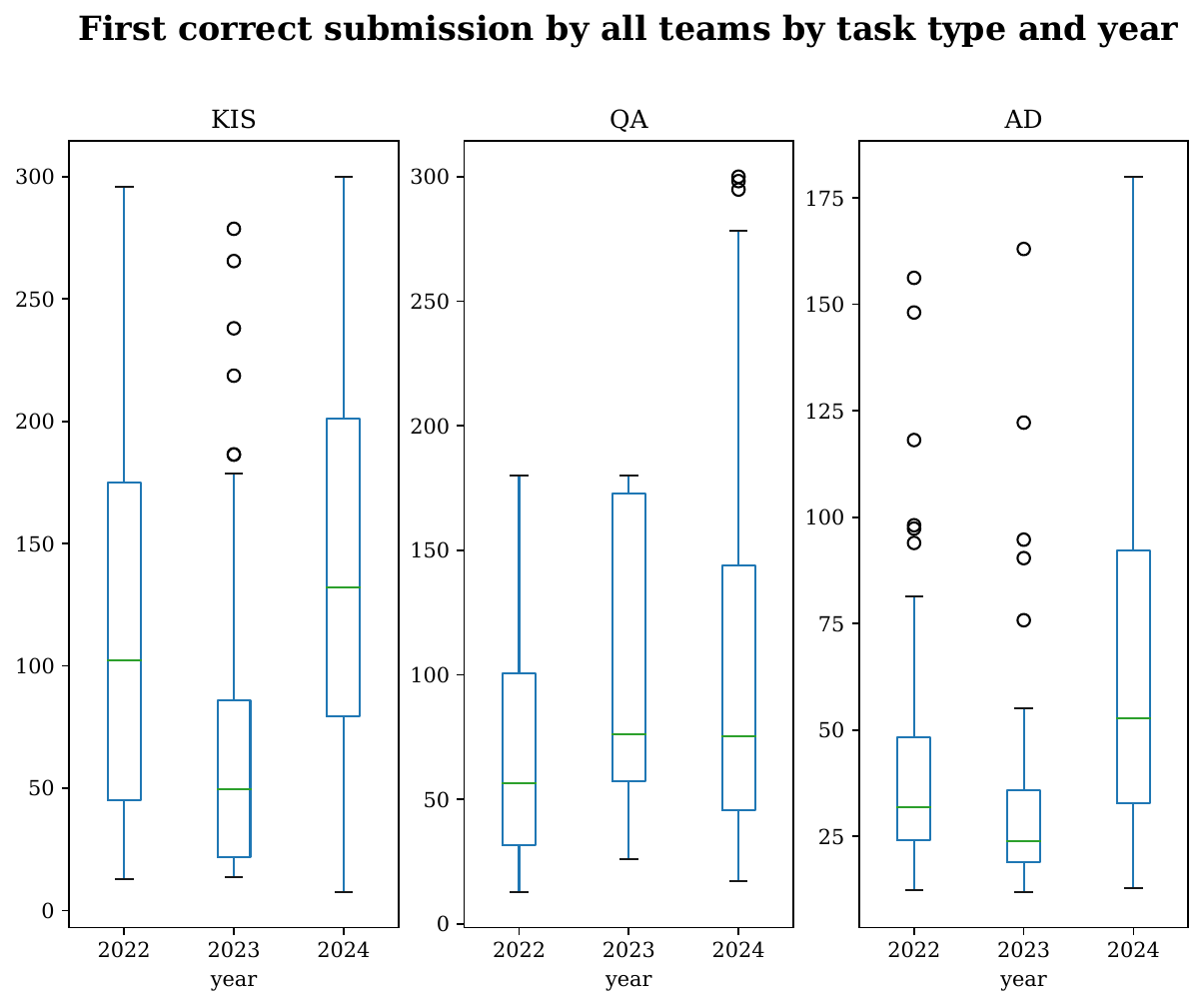}{Time to first correct submission by all teams in LSC'22, '23, and '24.\label{fig:first_correct_submission_by_all_teams_by_task_type_and_year}}

\subsubsection{Scores of Regular Teams by Task Type and Year}
\new{Next, we combine results at the team level to examine how scores changed over time for recurring participants.}
\Figure[h][width=0.99\columnwidth]{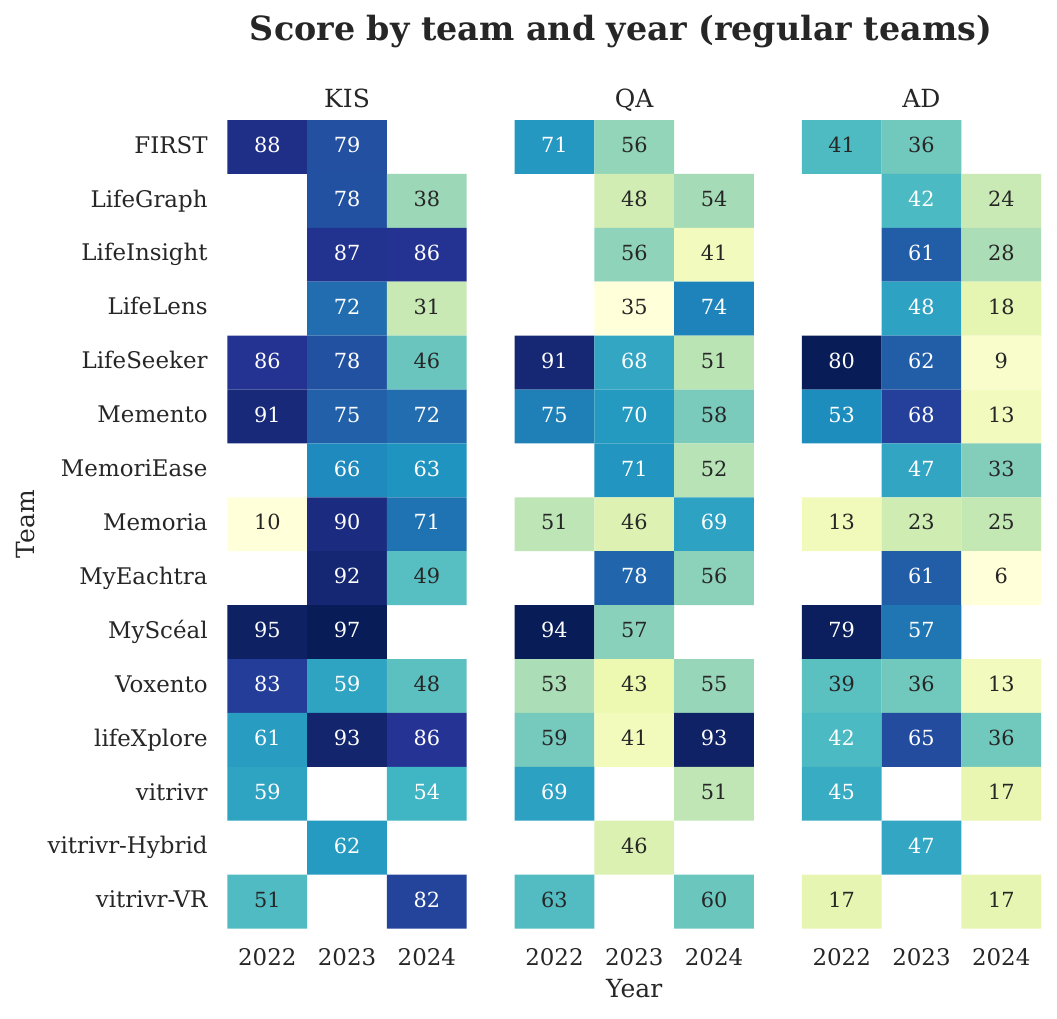}{Score of regular teams---those that participated in more than one year---in LSC'22, '23, and '24.\label{fig:score_by_team_and_year_regular}}
Figure~\ref{fig:score_by_team_and_year_regular} presents the scores of regular teams—those who participated in at least two editions—across KIS, QA, and Ad-hoc tasks from 2022 to 2024. The heatmap highlights year-on-year trends in performance. It is also important to note that the systems are changing every year to incorporate new approaches and techniques, which can impact performance. This analysis focuses on regular teams to understand if they have adapted and improved over time.

In KIS tasks, top-performing teams like MyScéal, LifeInsight, and LifeSeeker maintained consistently high scores across multiple years. However, LifeLens and LifeGraph saw significant score declines in 2024, potentially due to system changes or difficulties adapting to new task types. MEMORIA showed a notable improvement, increasing its score from 10 in 2022 to 90 in 2023, maintaining a strong performance in 2024. This suggests that substantial revisions were made to their system. In contrast, teams like MyEachtra and vitrir experienced steep declines in 2024, indicating that their new approaches may not have been as effective.

For QA tasks, MyScéal and LifeSeeker continued to perform strongly, while LifeLens significantly improved its score from 35 in 2023 to 74 in 2024, reflecting effective system updates. Conversely, LifeInsight faced a decline, indicating potential issues with adapting to the evolving QA task format. Teams such as FIRST and MemoriEase displayed year-to-year fluctuations, while vitrivr-VR and MEMORIA demonstrated more consistent performance.

For Ad-hoc tasks, MyScéal, LifeSeeker, and LIFEXPLORE achieved high scores across multiple years, indicating robust performance in open-ended tasks. However, MEMORIA and Memento struggled throughout, despite some year-on-year improvement. For example, MEMORIA increased its score from 13 in 2022 to 25 in 2024, but this remained below the top scores. The spread of scores across teams was widest for Ad-hoc tasks, reflecting the challenge of achieving high accuracy in unrestricted submission formats.

Overall, regular teams exhibited a mix of trends, with some improving significantly over time and others struggling to maintain consistency. It is also important to note that the systems are changing every year to incorporate new approaches and techniques, which can impact performance. The variability in scores across tasks and years highlights the complexity of lifelog retrieval and the need for teams to continuously adapt to remain competitive.
\new{These score patterns over time motivate a closer look at \emph{query difficulty}, which may help explain several of the fluctuations noted above.}

\subsection{Query Difficulty Analysis}\label{sec:query-difficulty}
\new{We now examine query characteristics to explain when and why systems (and instances) succeed or fail.}

\subsubsection{Number of Groundtruth Images}
\Figure[t!][width=0.99\columnwidth]{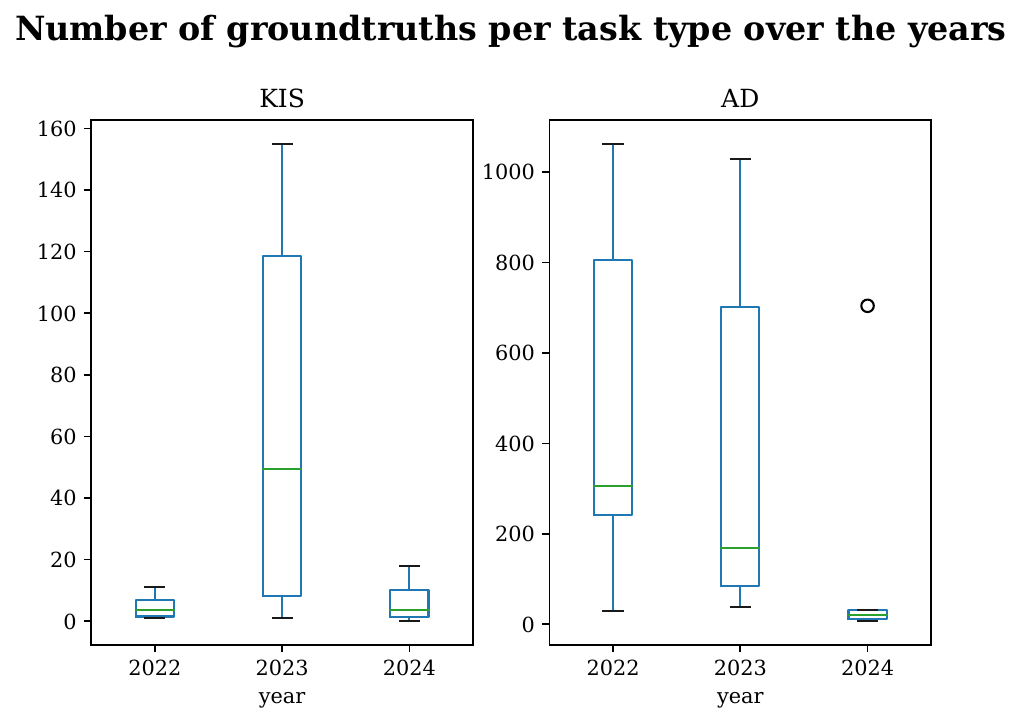}{Distribution of the number of groundtruth images in Ad-hoc and KIS tasks in LSC'22, '23, and '24.\label{fig:num_groundtruths}}

\Figure[t!][width=0.99\textwidth]{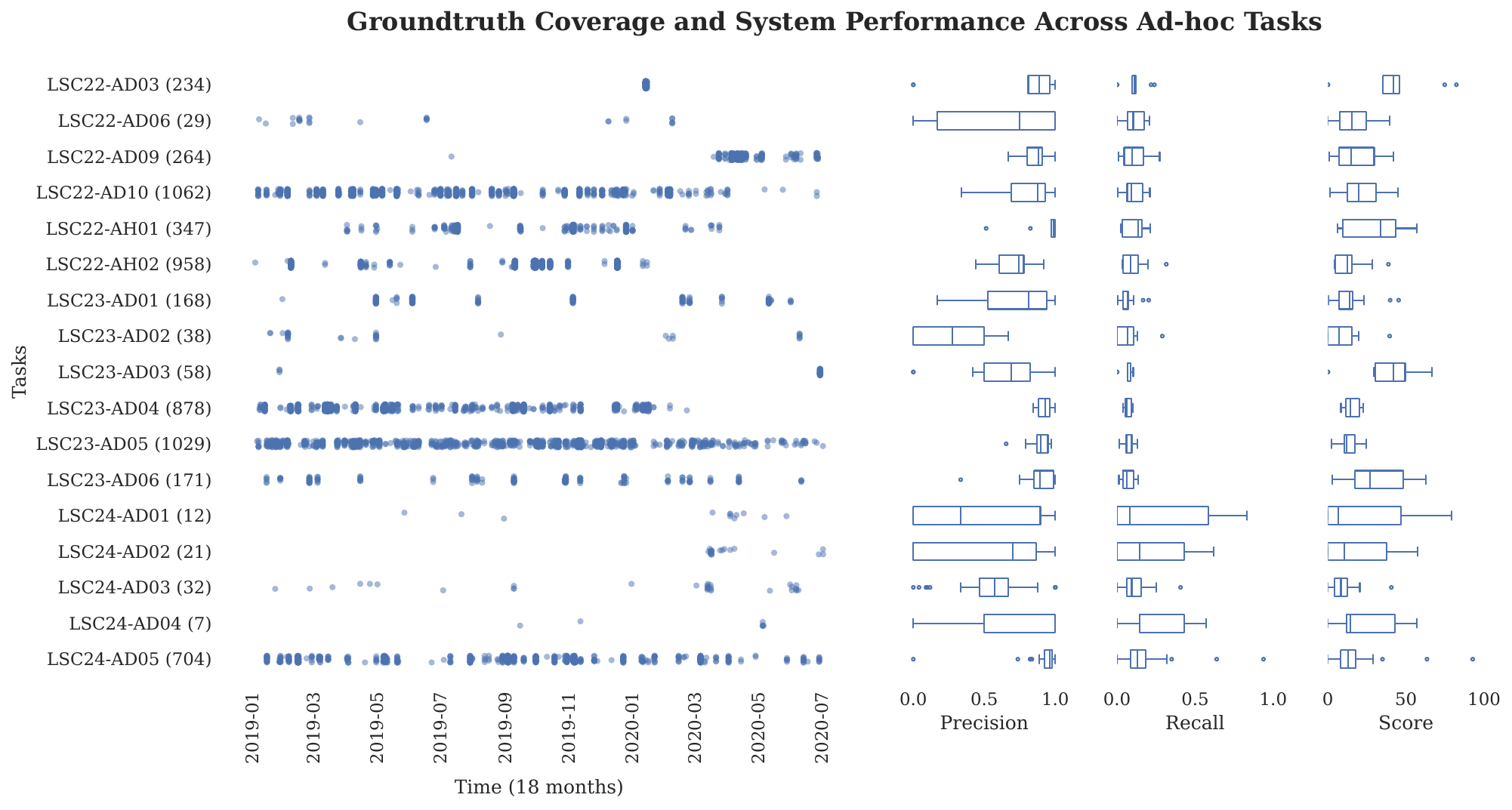}{Groundtruth coverage, precision, recall and score for Ad-hoc tasks.\label{fig:groundtruth_precision_score}}

Figure~\ref{fig:num_groundtruths} visualises the distribution of the number of groundtruth images associated with each task for KIS and ad-hoc retrieval tasks across the years 2022, 2023, and 2024. Groundtruth images are reference images that are considered correct for a specific query. We do not consider QA tasks in this analysis, as the number of groundtruth images does not apply to text-based answers.

In KIS tasks, the number of groundtruth images per query remained consistently low in 2022 and 2024, with the median staying close to a small value (under 10). However, in 2023, the distribution shows a significant increase in variability, with some queries having up to around 140--150 groundtruth images. This spike indicates that certain queries in 2023 were associated with longer events from the dataset, making the tasks less about pinpointing a unique image and more about identifying one from a broader set. This may reflect a subtle variation like the KIS topics that year, with some covering longer time spans than usual, possibly affecting how participants approached the topics. By 2024, the distribution had narrowed again, suggesting a return to queries with more targeted and unique references.

For Ad-hoc tasks, the distribution of groundtruth images has consistently shown a wide range across all years, reflecting the open-ended nature of these queries. In both 2022 and 2023, the median number of groundtruth images was fairly stable, but the spread extended from very few reference images to over 1000 in some cases. In 2024, however, the range and median dropped substantially, indicating that the number of valid reference images per query was reduced. This change potentially made the task more challenging by narrowing the acceptable reference space, requiring systems to be more precise in their submissions, similar to the KIS task. The reason for this shift is unclear, but it may be due to the depletion of novel Ad-hoc queries, as many queries had already been used in previous years. This resource limitation may have impacted the open-ended nature of the task.

The trends in the number of groundtruth images can help explain some of the system performance results discussed earlier. For KIS tasks, the spike in 2023 groundtruth counts aligns with the higher number of correct submissions and increased precision seen in 2023. Having more reference images per query likely increased the chances of correct matches, making it easier for systems to provide valid answers. This contrasts with 2024, where the reduced number of groundtruth images contributed to a decline in correct submissions and precision. For Ad-hoc tasks, the narrowing of the distribution in 2024 corresponds to the higher performance variability observed among system instances. Queries with fewer groundtruth images require more precise retrieval, which may have exacerbated performance disparities between well-optimised and poorly configured systems. Additionally, if the remaining queries were inherently more difficult or ambiguous, this could explain the increased difficulty in achieving high scores despite improved system tuning.
\new{Having established how groundtruth coverage shifted across years, we next consider the temporal \emph{distribution} of those groundtruths and their link to precision/recall.}

\subsubsection{Groundtruth Coverage and System Performance Across Ad-hoc Tasks}
Figure~\ref{fig:groundtruth_precision_score} presents the temporal distribution of groundtruth images for each Ad-hoc task alongside system performance metrics: precision, recall, and score. The scatter plot shows timestamp distributions, while the box plots compare retrieval performance across tasks.

A key observation from the scatter plot is the variability in the temporal spread of groundtruth images across tasks and years. In 2022 and 2023, tasks were more evenly distributed across timestamps, resulting in more consistent patterns of performance. However, in 2024, the temporal distribution of groundtruth images was more clustered, with some tasks having a high concentration of reference images around specific timestamps. This clustering may have contributed to the increased performance variability observed in 2024. This raises concerns about the nature of the tasks and their differentiation from KIS queries, which are typically more targeted.

To understand the impact of groundtruth distribution on system performance, we conducted a correlation analysis between the temporal spread of groundtruth images and retrieval metrics. To quantify the degree of `well-distributed' groundtruth images, we calculated the Gini coefficient for each task, where a lower value indicates a more even spread of timestamps and a higher value indicates clusters in specific time spots.

The results show some negative correlation between the Gini coefficient and recall ($r = -0.15$, $p = 0.13$) and score ($r = -0.24, p = 0.05$), suggesting that unevenly distributed groundtruth images may hinder retrieval performance, even though the effect is relatively weak. This pattern aligns with the scatter plot observations, where tasks with more evenly spread reference images generally performed better than those with clustered timestamps. The correlation between the Gini coefficient and precision was near zero ($r = -0.03$, $p = 0.24$), indicating that timestamp distribution appears to affect recall and coverage more than precision.

The number of groundtruth images per task showed a weak negative correlation with recall ($r = -0.17$, $p = 0.097$) and score ($r = -0.078$, $p = 0.20$), suggesting that retrieval effectiveness may depend more on system design and query formulation than on the sheer availability of reference images.

The correlation analysis also revealed that recall remains the strongest predictor of high scores ($r = 0.75$, $p = 0.025$), reinforcing the importance of retrieving a larger number of correct results, even at the expense of including some incorrect ones.
Precision showed a moderate but statistically non-significant correlation with scores ($r = 0.49$, $p = 0.42$), so its contribution to overall performance remains unclear.
As observed in the box plots, recall scores were generally low, while precision scores were more evenly distributed. This implies that many systems struggled to retrieve enough relevant images, even if they maintained reasonable accuracy when they did.

These findings emphasise the need for future lifelog retrieval systems to adapt their strategies based on both the quantity and temporal distribution of reference images, ensuring more robust and effective retrieval outcomes.
\new{In summary, query difficulty, through both groundtruth size and temporal spread, helps explain the correctness and timing patterns observed earlier. We now look at how these factors appear in LSC’24 specifically.}

\subsection{LSC'24: Performance Analysis}\label{sec:performance}
In this section, we take a closer look at the performance of systems in the LSC'24. Novice tasks were excluded from this analysis due to wide variability in system performance. In each figure, the teams are ordered by their overall score in the respective task type (KIS, QA, Ad-hoc), which can be seen in the leaderboard in Figure~\ref{fig:overall}. We follow the same structure as in Section~\ref{sec:comparison}, focusing on the correctness of submissions, time to the first correct submission, and overall scores.
\new{This follows the same sequence as before, allowing direct comparison between cross-year trends and 2024-specific results.}

\Figure[h][width=0.99\columnwidth]{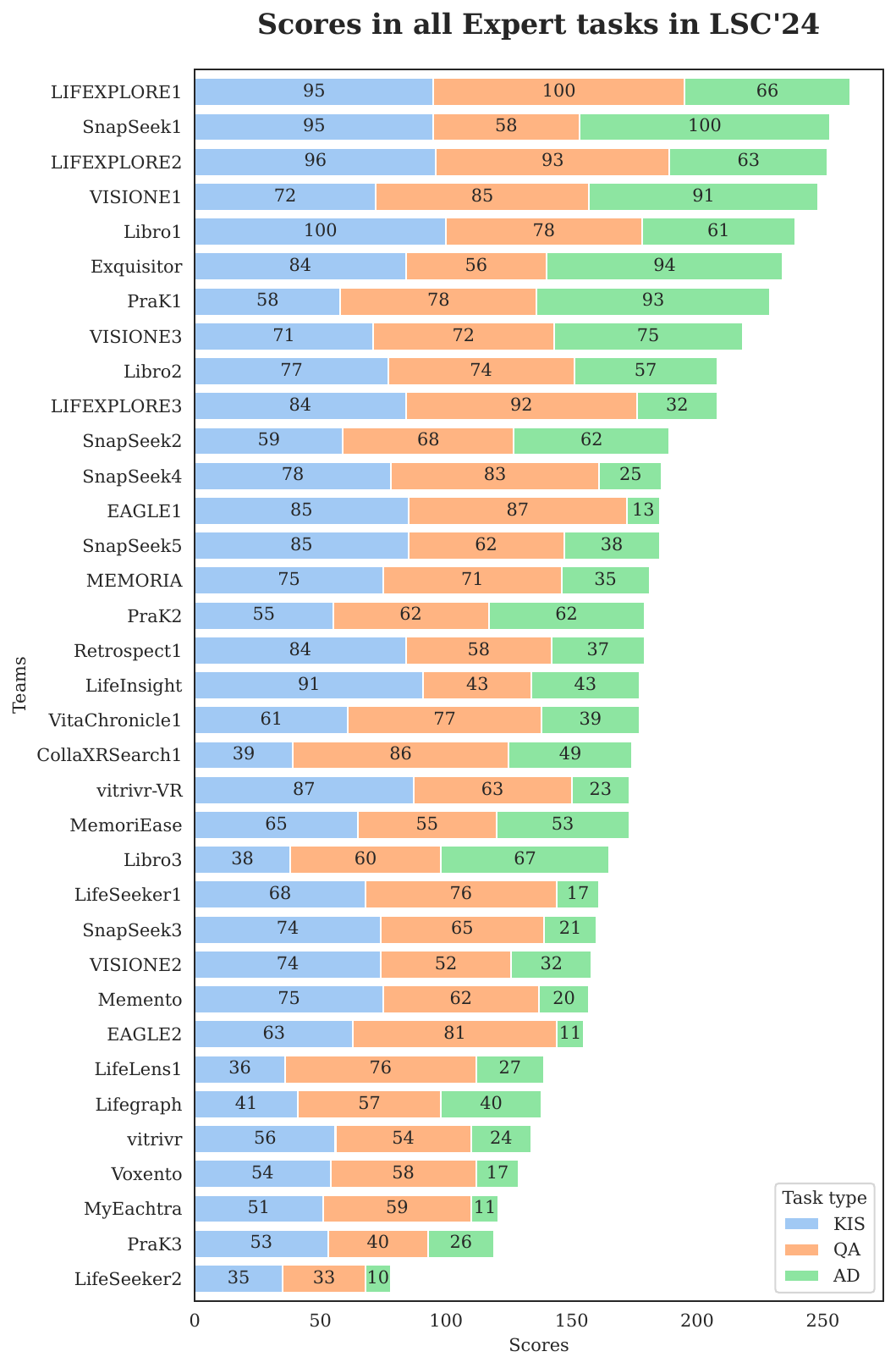}{Overall performance in LSC'24.\label{fig:overall}}

\subsubsection{Correctness of Submissions}
We observed a direct correlation between the number of correct submissions and overall scores in KIS and QA tasks, as shown in Figure~\ref{fig:correct_incorrect}. Teams that submitted more correct answers tended to achieve higher scores. Lower-scoring teams often made multiple incorrect attempts, which reduced their overall performance. In ad-hoc tasks, since the number of groundtruth images varied widely across queries, the number of correct submissions was less indicative of overall performance as the scores were normalised per task. Therefore, we focused on precision and recall per task to evaluate system performance in ad-hoc tasks, as shown in Figure~\ref{fig:precision-recall}. In this graph, we observed a stronger correlation between recall and overall scores, compared to precision. Even though penalties were applied for incorrect submissions, recall played a more significant role in determining the final score as described in the evaluation methodology. Nevertheless, the top-performing teams still maintained high precision, indicating a balanced approach.
\new{Next, we examine whether timing can help explain the variation observed in correctness and precision.}

\Figure[t!][width=1.0\textwidth]{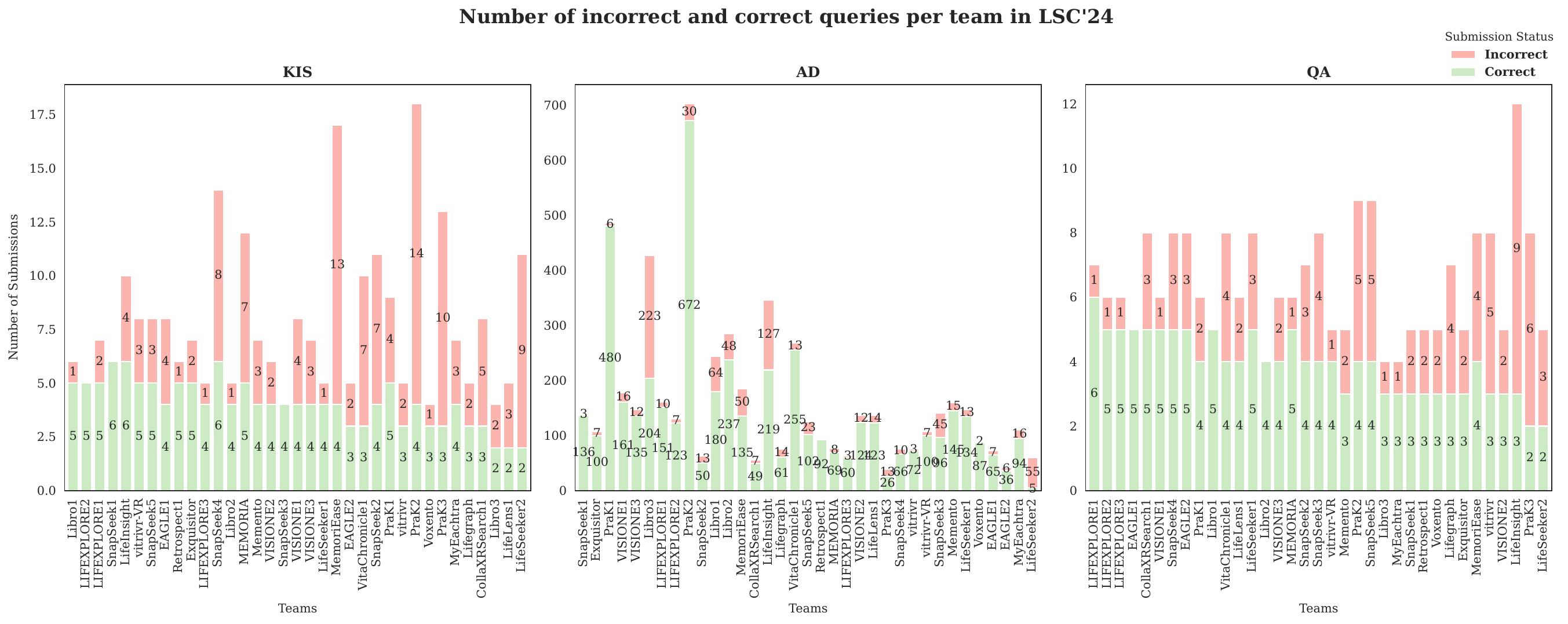}{Number of correct and incorrect submissions in LSC'24.\label{fig:correct_incorrect}}
\Figure[h][width=0.99\columnwidth]{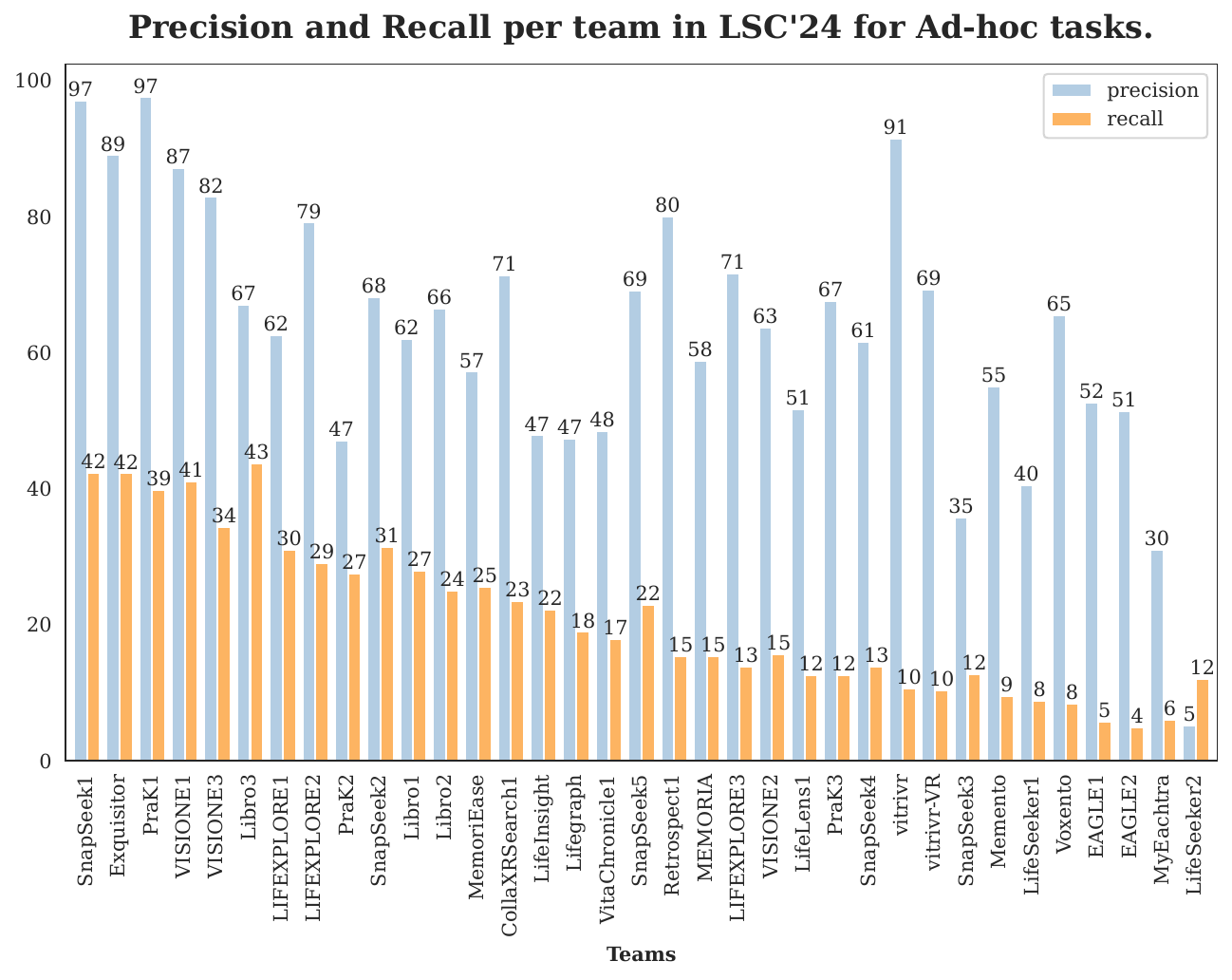}{Precision and recall for Ad-hoc tasks in LSC'24.\label{fig:precision-recall}}

\subsubsection{Time to Correct Submission}
Figure~\ref{fig:time} shows the time taken by teams to submit their first correct answer in LSC'24. In KIS tasks, most teams submitted correct answers within 100--200 seconds, with a median time (of all correct submissions) of 132 seconds. There seems to be no clear correlation between the time taken to submit the first correct answer and the overall score. In QA tasks, the time to correct the submission was generally shorter. Going down the leaderboard, the time taken to submit the first correct answer increased, then decreased again for the lowest-scoring teams. This suggests that the lowest-scoring teams struggled to submit a correct answer for many queries, but when they did, they did so quickly. The middle-scoring teams took longer to submit their first correct answer; however, they managed to submit more correct answers overall, leading to higher scores. A more consistent pattern is observed in Ad-hoc tasks, where the time to the first correct submission is generally inversely proportional to the overall score. Teams that took longer to submit the first correct answer tended to have lower scores, while those that submitted correct answers quickly achieved higher scores. This indicates that efficient query formulation and retrieval strategies are crucial for success in Ad-hoc tasks.
\Figure[h][width=1.0\textwidth]{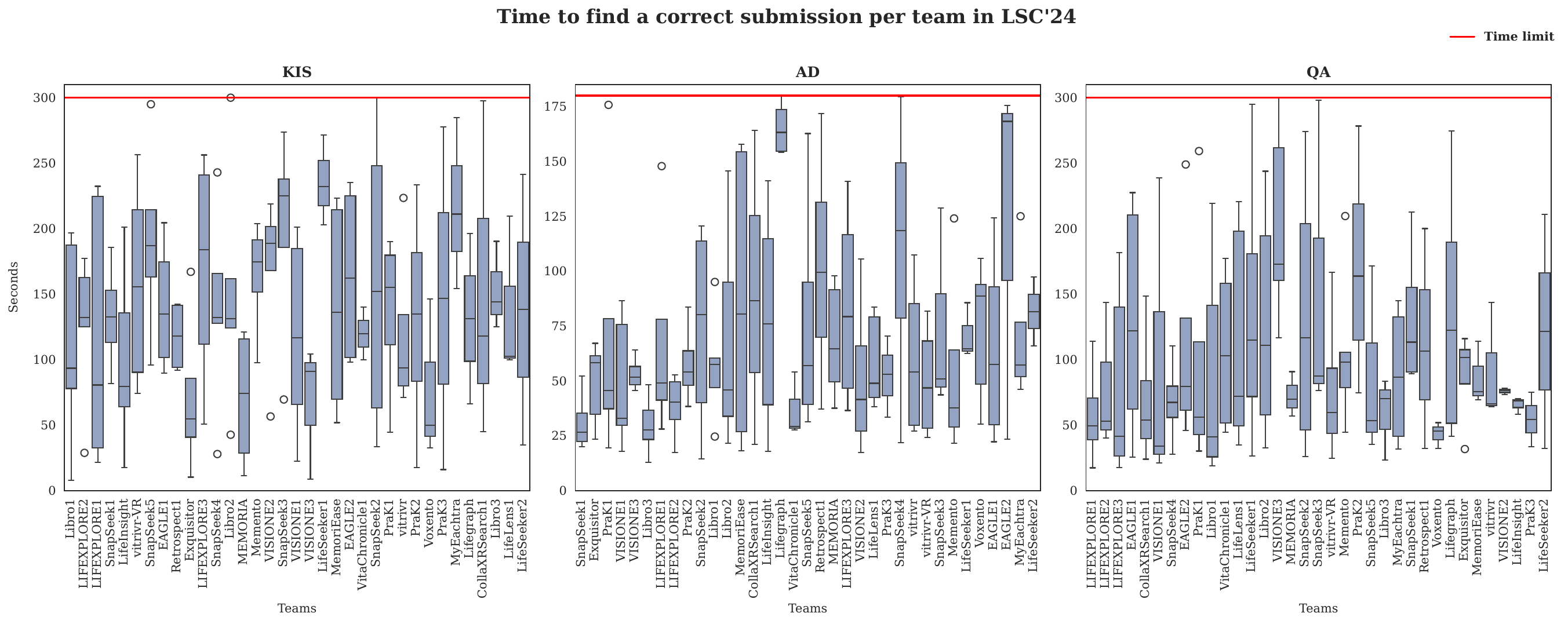}{Time to first correct submission in LSC 2024.\label{fig:time}}
\new{We now turn to score distributions to characterise consistency across tasks and teams.}

\subsubsection{Score Distribution}
In order to understand more about the performance of teams in LSC'24, we analysed the distribution of scores across each individual task in Figure~\ref{fig:performance}. The heatmap reveals clear distinctions in performance, highlighting top teams, task-specific challenges, and consistency across different retrieval methods.

\Figure[ht!][width=1.0\textwidth]{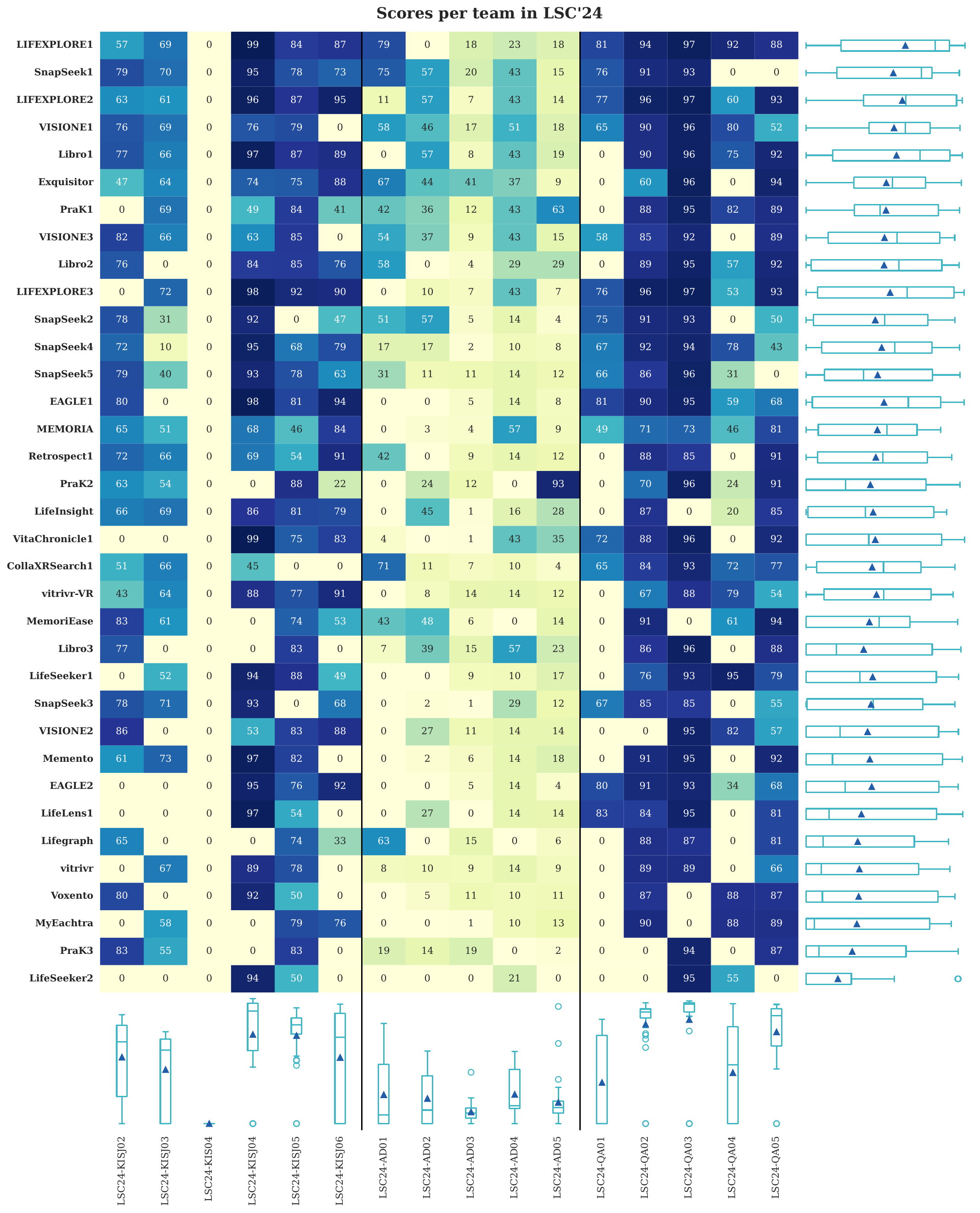}{Performance of teams in LSC'24.\label{fig:performance}}
Box plot analysis provides insights into task difficulty and performance disparities across different retrieval types. KIS tasks generally had high and stable scores, reinforcing that these tasks were easier to solve. In contrast, ad-hoc tasks exhibited the lowest scores, and QA tasks showed the most variability, with extreme differences between easy and complex queries. This variation highlights how different retrieval strategies impact performance, with structured searches benefiting from well-defined queries while open-ended searches introduce greater challenges.

A key observation from the results is the difference in consistency versus variability across teams. LIFEXPLORE1, SnapSeek1, and VISIONE1 consistently maintained high scores across all tasks, demonstrating well-rounded retrieval capabilities. In contrast, lower-performing teams showed greater fluctuations, indicating difficulties in handling certain types of queries—especially those requiring semantic understanding, query expansion, or temporal reasoning. Notably, teams that operated multiple user groups, such as PraK1, PraK2, and PraK3, exhibited drastic performance differences despite using similar retrieval systems. PraK2 performed exceptionally well, scoring 93 in LSC24-AD05, while PraK1 and PraK3 struggled in most tasks. This suggests that user variability plays an important role in retrieval performance, potentially due to differences in search strategies, system familiarity, or individual decision-making approaches. Similar inconsistencies were observed among SnapSeek variants, where some instances performed reliably, while others showed notable fluctuations.

Performance differences also varied across task types. KIS tasks were generally the easiest, with most teams achieving high scores. SnapSeek1 and LIFEXPLORE1 consistently outperformed in this category. Box plot analysis reveals that for five out of six KIS tasks, the average score remained above 42, while the median score exceeded 58, indicating that most retrieval models handled these queries effectively. However, KIS04 was an exception, with no teams successfully retrieving the correct answer.

Upon closer inspection, KIS04 presents an anomaly, as all teams scored zero, which is highly unusual for KIS tasks. The query hints included: \textit{`It was as if the meeting took place on the Starship Enterprise. I remember the wall had yellow lighted shapes and designs, with a yellow lighting on the ceiling. There were 3 or 4 people in the meeting room with me. It was a formal boardroom. Afterwards, I got a taxi back to my hotel in Bangkok.'} This query was conceptually vague, relying on subjective interpretation rather than concrete object-based details. The metaphorical reference to the Starship Enterprise may have confused both users and retrieval models, making it difficult to identify relevant images. By the time specific visual details were provided (e.g., yellow lighting, boardroom, people in the room), users may have already formed a completely different mental image, leading to misinterpretations. This anomaly underscores the importance of clear, structured query design to ensure that retrieval tasks remain solvable. \new{This also illustrates a broader limitation of current multimodal retrieval: while embedding models align literal or concrete descriptions well, they still struggle with figurative, cultural, or metaphorical language. Bridging this `semantic gap' between abstract phrasing and visual grounding remains an open research challenge, demanding hybrid models that integrate world knowledge and commonsense reasoning with visual grounding.
}

QA tasks exhibited the highest variation in difficulty. While some questions were straightforward, most teams scoring close to 100, others required complex reasoning or multi-step retrieval strategies, resulting in nearly half of the teams scoring zero. This sharp contrast suggests that QA tasks require a more refined balance between difficulty levels, ensuring that some level of inference is necessary without making questions entirely unsolvable.

Ad-hoc tasks posed the greatest challenge, with notably lower scores across teams. The average scores for these tasks were under 23, while the median scores remained below 14. While LIFEXPLORE1, VISIONE1, and SnapSeek1 maintained competitive results, many teams struggled to achieve meaningful retrieval results, often scoring near zero in multiple tasks. One notable case was LSC24-AD05, where PraK2 achieved 93, while PraK1 scored 63, creating a large disparity that made other teams appear weaker due to normalisation effects in the pooled evaluation metric. This suggests that some ad-hoc queries may be highly system-dependent, favouring specific retrieval models over others.

Lower-performing teams should refine their strategies for open-ended ad-hoc tasks to improve future performance, as these exhibited the greatest variability and lowest overall scores. Even the top teams could benefit from improving consistency across different retrieval types, ensuring robust performance not only in structured KIS queries but also in more semantically complex ad-hoc and QA tasks. Addressing these challenges will be crucial for advancing lifelog retrieval systems and improving their adaptability to diverse search scenarios.
\new{In sum, LSC'24 mirrors the cross-year story: high KIS solvability, volatile QA difficulty, and ad-hoc sensitivity to instance/user effects and query design.}

\subsection{Insights on Techniques and System Performance}\label{sec:technique-performance}
\new{Finally, we synthesise the above results into technique-level insights, tying method choices back to the correctness, timing, and score patterns.}
This section summarises key insights regarding how the retrieval techniques employed by different teams influenced their overall system performance across the LSC tasks. Even though the performance of systems varied widely, the following trends can be observed:

\textbf{Embedding-based Retrieval.} Embedding-based retrieval methods showed a strong positive correlation with the system scores in LSC'22 in all task types.\footnote{Spearman's correlation coefficient: AD: 0.48, $p<10^{-3}$; KIS: 0.40, $p=0.00$; QA: 0.27, $p=0.01$} However, as most teams adopted embedding-based retrieval methods in 2023 and 2024, there was no significant correlation between embedding usage and system performance, suggesting that other factors played a more critical role in determining system success in these years. For example, the choice of embedding model, tuning parameters, and system design may have influenced performance more than the mere adoption of embeddings. The saturation effect of embedding-based retrieval methods highlights the need for teams to explore alternative approaches to maintain a competitive edge in future LSC editions.

\textbf{Large Language Models and Conversational Search.}
Although LLM-driven conversational retrieval methods have been increasingly adopted, their impact on system performance remains unclear. The top-performing QA systems in recent LSC editions (e.g., LIFEXPLORE, LifeLens, MEMORIA in 2024, and MyEachtra/MyScéal in earlier years) were not explicitly reliant on conversational LLM techniques. Instead, their successes suggest that effective handling of queries, robust semantic retrieval strategies, and system refinement played a more critical role in achieving high QA scores.

\textbf{UI/UX Design and Retrieval Efficiency.}
Teams like Libro clearly benefited from streamlined UI/UX designs, as demonstrated by their rapid retrieval times, especially in KIS tasks. However, minimalism alone, as exemplified by LifeLens, did not necessarily translate to faster retrieval. In contrast, high-scoring teams such as LIFEXPLORE and SnapSeek balanced sophisticated retrieval methods with effective UI/UX, suggesting a synergistic rather than purely causal relationship. Therefore, while effective UI/UX design can enhance retrieval efficiency and performance, it is not a standalone predictor of success.

\textbf{Multimodal and Collaborative Approaches.} Innovative approaches like collaborative retrieval (\textit{CollaXRSearch}) and eye-tracking integration (\textit{EAGLE}) introduced potential advantages. However, their complexity may introduce performance variability, as seen in the uneven performance across task types. These approaches require further refinement to achieve consistent effectiveness across diverse queries.

\textbf{Impact of Users} Teams with multiple instances (e.g., \textit{SnapSeek}, \textit{PraK}, \textit{VISIONE}) experienced significant performance variability. The dramatic differences in retrieval outcomes between system instances highlight the critical role of user familiarity and effective UI/UX design in achieving high performance. In addition, systems that shared the same core retrieval technology but differed in interface design (e.g, LifeLens using LifeSeeker's engine), deployment configuration, or user guidance also showed varying levels of effectiveness. These observations reflect how the LSC evaluates systems in an end-to-end manner, where both retrieval quality and user interaction contribute to final performance.

Ultimately, the most significant insight from this analysis is that no single retrieval technique or design choice conclusively predicted system performance. Instead, successful lifelog retrieval systems seem to depend upon careful integration, tuning, and balanced design choices rather than any isolated retrieval method or technology. Additionally, deploying multiple instances of the same system in expert-track evaluations demands careful management of user familiarity. Such multi-instance designs might better suit the novice track in future LSC editions, where variability due to user experience can be more meaningfully assessed.
\new{Taken together, these points unify the metric trends and query-difficulty results, concluding the section’s analysis.}

\section{Conclusion}\label{sec:conclusion}
Over the past three years, LSC has witnessed significant advances in lifelog retrieval techniques, user interfaces, and system performance. The 2024 edition marked a substantial shift towards integrating LLMs, multimodal retrieval, and enhanced user interfaces, building upon the embedding-based methods that became dominant in LSC'22 and LSC'23.

\subsection{Technical Advancements}
Embedding-based retrieval has emerged as a standard technique, with models such as CLIP, BLIP, and OpenCLIP playing a central role in text-image search. The adoption of weighted ensemble approaches, as seen in Memento 4.0 and LifeInsight 2.0, has further refined retrieval accuracy. The 2024 edition also saw the rise of LLM-powered systems, such as MemoriEase and Memento 4.0, leveraging Retrieval-Augmented Generation (RAG) to provide more interactive and context-aware lifelog search experiences. Voxento-Pro integrated OpenAI's Assistant API, demonstrating the transition towards voice-based conversational retrieval. Other systems, such as MyEachtraX and LifeGraph 4, incorporated multimodal LLMs, combining text and visual cues to improve retrieval performance. The expansion of LLMs in lifelog search underscores their ability to refine query understanding, enhance response generation, and facilitate more interactive search experiences.

User interface improvements have been crucial in enhancing usability, with systems like LifeLens 2.0, VitaChronicle, and Retrospect focusing on minimalist, task-oriented designs, while CollaXRSearch introduced a shared virtual workspace for collaborative search. Additionally, vitrivr-VR advanced VR-based interaction with spatiotemporal queries. Beyond these developments, novel search paradigms such as eye-tracking-based retrieval, demonstrated by EAGLE, and collaborative VR search, as implemented in CollaXRSearch, have expanded the boundaries of lifelog exploration, improving precision and user experience.

\subsection{System Performance and Key Observations}
The analysis of LSC'22, '23, and '24 highlights key trends in system performance. KIS and QA tasks consistently yielded high precision and shorter response times, whereas Ad-hoc retrieval tasks showed greater variability due to their open-ended nature and broader temporal groundtruth distribution. The top-performing teams effectively balanced precision, recall, and efficient submission strategies, while lower-performing teams often struggled with precision or relied on brute-force submissions.

Scores in KIS and QA tasks improved from 2022 to 2023 but became more variable in 2024 due to the introduction of multiple system instances with varied users or configurations. Correct submissions were closely correlated with the number of available groundtruth images, with 2023’s KIS tasks benefiting from a higher count, while 2024’s reduced number of groundtruth images increased difficulty.

Performance varied notably across task types. Ad-hoc tasks remained the most challenging, displaying wide variance in team performance on precision and recall. Systems such as LIFEXPLORE, SnapSeek, and VISIONE demonstrated consistently high scores, while others exhibited fluctuations due to differences in instances. Additionally, LLM-integrated systems like MemoriEase, Memento 4.0, and Voxento-Pro excelled in QA tasks, showcasing the effectiveness of conversational search and RAG techniques in refining retrieval precision. The failure of KIS04, however, underscored how abstract or ambiguous prompts, while arguably closer to natural memory recall, can be particularly difficult for current retrieval systems. This highlights the ongoing challenge in KIS task design: while realistic, memory-like queries are desirable, they must still be feasible for systems operating without prior user context. This case demonstrates the importance of calibrating ambiguity and providing enough grounding to enable meaningful evaluation.
However, as the LSC enters its eighth year, it is appropriate for tasks to become more challenging to test the evolving capabilities of state-of-the-art systems. Realistic, memory-like queries—often imprecise or fuzzy—better reflect real-world information needs in lifelog retrieval.

\subsection{Limitations}
This paper provides an overview of system performance in LSC 2024, but several limitations must be acknowledged. \new{First, while privacy and ethics are central to lifelogging in general, this work focuses on retrieval performance under an already ethically approved dataset.}

Second, the analysis is based on a dataset collected from a single individual, which limits its demographic, cultural, and behavioural representativeness. While it enables longitudinal analysis and consistent annotation, the findings may not generalise to more diverse lifelogging contexts. However, we would like to note that the lifelogger travelled to many countries across multiple continents, thereby capturing a wide variety of cultural contexts. The unique strength of the dataset lies in the continuous, long-term data collection, offering rare insights into behavioural patterns over time.

Third, we excluded novice systems from quantitative analysis due to their inconsistency and incompleteness, which may limit insights into first-time usability. Performance variations among systems with shared cores (e.g., SnapSeek variants) further reflect the influence of user factors outside the scope of this analysis.

Moreover, while all systems within a given LSC edition are evaluated on the same set of tasks under identical conditions, ensuring fairness within each year, comparisons across years are more nuanced. Variability in task difficulty, especially in subjective or open-ended tasks, may influence performance trends. Although we mitigate this by averaging scores across multiple tasks and reporting per-task-type performance, a fully consistent difficulty calibration across years remains an open challenge. Achieving this would likely require re-running historical systems on current-year tasks, ideally with the same users and interfaces-something not currently feasible in the live competitive format.

Additionally, Gini was used to reflect temporal imbalance due to its interpretability and simplicity. We acknowledge it is a simplification, and alternatives such as entropy or burstiness could offer richer descriptions of temporal structure. However, given space constraints and our focus on system behaviour rather than metric development, we leave deeper exploration to future work.

Finally, this review does not address computational scalability, accessibility considerations (e.g., eye-tracking), or hallucinations and privacy risks introduced by generative components such as LLMs. These aspects, while important for real-world deployment, fall outside the scope of this paper, which is constrained by space and focused on analysing submitted interactive system performances within the competition context. We consider these critical directions for future work and recommend targeted studies to investigate them in depth.

\subsection{Future Directions}
Building on our longitudinal analysis and query difficulty findings, we outline future research directions that address observed system trends, evaluation limitations, and emerging interaction challenges in lifelog retrieval.
\begin{itemize}
	\item \textbf{Refinement of Conversational Search}: While LLMs have improved interactive retrieval, there remains room for enhancing temporal and situational context awareness and improving response accuracy. Conversational search may also support iterative refinement of ambiguous queries through follow-up prompts or clarifications, complementing foundational retrieval efforts. Future iterations require investigation of fine-tuning models on lifelog-specific datasets to better capture personal, temporal, and activity-related semantics.
    
	\item \textbf{Optimised Ad-hoc Query Handling}: Addressing performance disparities in Ad-hoc retrieval requires improved search strategies, specifically through automated query expansion (using generative AI, etc.) and interactive negative relevance feedback that allows users to quickly filter out visually similar but contextually irrelevant false positives.
	% \item \textbf{Enhanced Personalisation and Adaptability}: Systems could integrate user-specific preferences and search behaviours to provide more tailored retrieval experiences. This could involve hybrid approaches that combine embeddings, knowledge graphs, and user feedback loops.
	\item \textbf{Improved Evaluation Metrics}: Given the increasing complexity of lifelog retrieval tasks, more nuanced evaluation metrics that account for query complexity, retrieval diversity, and the cognitive or interaction cost could provide better performance insights. Specifically, we suggest adopting graded relevance and Interaction-Cost-to-Success to measure search efficiency alongside existing metrics.
	% \item \textbf{Multimodal Fusion and Advanced Interfaces}: Integrating text, voice, and image-based inputs can enable users to express their queries more intuitively. Virtual and augmented reality interfaces may further enhance user interaction by providing immersive environments where users can “relive” events through reconstructed lifelog data. However, practical challenges remain in adopting these interfaces widely.
	% \item \textbf{Event Summarisation and Automated Narratives}: As lifelogs capture increasingly granular details, summarisation tools that generate coherent narratives from raw data will become essential. These tools could help users quickly review their daily activities, significant milestones, or recurring patterns, improving accessibility and usability.
	% \item \textbf{Collaborative Search Environments}: Collaboration features, such as shared lifelog workspaces and annotation tools, can support joint exploration of lifelog data for purposes such as health monitoring, caregiving, or memory sharing. Developing secure and user-friendly collaborative search environments will enable users to co-navigate their lifelog collections with trusted individuals while addressing privacy and security concerns.
        \item \new{\textbf{Controlled Studies with Novice Users} While expert performance provides a controlled view of system capabilities, future work should systematically examine user variability under realistic conditions. Controlled studies could recruit both novices and semi-experienced users to assess how familiarity, interface complexity, and retrieval transparency affect task efficiency and satisfaction. Mixed-method evaluations—combining task analytics with think-aloud or post-task reflection—would enable quantifying learning effects and uncovering usability barriers specific to first-time lifelog users.}

\end{itemize}

Especially given our query difficulty analysis in Section~\ref{sec:query-difficulty}, future work could explore optional guidance for lifeloggers during query formulation, with the aim of balancing authenticity of memory recall and retrievability. Some initial pointers could include:

\begin{itemize}
	\item Queries should aim to reflect realistic memory recall but avoid phrasing that is too vague or abstract, especially if no strong visual cues are present. Instead, prompts should strike a balance between authenticity and retrievability.
	\item For KIS tasks, while the memory context can involve longer events, the specific memory being recalled should ideally refer to a bounded, single-instance experience rather than an extended or recurring activity. This helps avoid overly broad groundtruth sets. Alternatively, KIS tasks could be split into two categories: `specific event' and `abstract/ambiguous memory'. This would acknowledge the diversity of real memory recall while allowing systems to be evaluated more meaningfully within each category.
	\item For Ad-hoc tasks, we recommend introducing a tool for visualising and monitoring the distribution of groundtruth images, both in terms of quantity and temporal spread. This would help organisers identify skewed or inconsistent task profiles before release.
	\item Also for Ad-hoc tasks, one promising direction is to move beyond scoring individual images and instead evaluate system outputs at the level of temporal segments. Many Ad-hoc queries implicitly refer to extended activities (e.g., `using chopsticks at a meal', `eating fast food'), and this approach would allow evaluation metrics to reflect how well systems retrieve relevant episodes, not just isolated images. The current image-level scoring allows for a potential shortcut: once a system retrieves a single image from a long groundtruth event, it can submit the surrounding frames to accumulate high scores, even if their overall retrieval strategy lacks precision or coverage. Segment-based scoring would reduce the impact of such strategies by emphasising temporal coherence and meaningful coverage. Moreover, adopting a segment-based approach better reflects realistic retrieval behaviour, as users tend to search for moments or episodes rather than individual frames. Segment-level metrics further assess a system’s ability to maintain temporal continuity across related scenes. Although this is not currently feasible due to the lack of a shared segmentation standard, we see clear motivation for shifting from image-level to segment-level metrics. We view the transition from image-level to segment-level evaluation not as an immediate solution, but as a longer-term research agenda requiring shared segmentation standards and new temporal evaluation metrics.
\end{itemize}

However, aligning evaluation methods with user needs requires more than technical validation. Future work should incorporate user studies or expert-in-the-loop processes to assess whether such metrics correspond to how users recall, search for, or interpret lifelog events. This may involve collecting qualitative feedback, modelling user intent, or running controlled studies to determine whether the proposed metrics reflect perceived relevance or retrieval satisfaction.

Finally, addressing dataset scarcity while preserving participant privacy remains a foundational challenge for lifelog research. Synthetic data generation (e.g., \cite{tan2023timelineqa}) and simulation-based lifelogging offer privacy-preserving alternatives. In parallel, repurposing existing egocentric video datasets, such as EPIC-Kitchens~\cite{Damen2022RESCALING}, Ego4D~\cite{DBLP:conf/cvpr/GraumanWBCFGH0L22}, or the recently released EgoLife~\cite{yang2025egolife} and CASTLE~\cite{rossetto2025castle}, provides a promising route for evaluating lifelog-style retrieval systems in naturalistic settings. While these datasets are not full lifelogs, task-specific annotations can adapt them for lifelogging benchmarks. We encourage future work to explore shared annotation efforts on such datasets and to develop protocols for synthetic, multi-perspective lifelog data collection.

\subsection{Final Thoughts}
Although the LSC takes place in a constrained, competitive environment, many of its task formulations closely mirror real-world lifelogging scenarios—such as recalling past events, locating personal media, or answering context-rich questions. This alignment suggests that the findings from LSC submissions are not only useful for benchmarking system performance but also transferable to the development of practical, user-facing lifelog retrieval tools.

Over the years, the LSC has showcased a diverse range of retrieval strategies, from advanced embedding-based search techniques to cutting-edge LLM-driven conversational interfaces.
Collectively, the competition entrants highlighted the growing sophistication of lifelog retrieval systems, as well as providing a reminder of the challenges that remain in making these systems robust, intuitive, and efficient. With continued innovation in AI, user interaction, and multimodal search, we are excited to see how the performance and ease-of-use of lifelog retrieval systems evolve in the coming years.

\section*{Author Contributions}
\noindent\textbf{Allie Tran:} Conceptualisation, Formal analysis, Investigation, Writing – Original Draft, Visualisation, Writing – Review \& Editing

\noindent\textbf{Werner Bailer:} Writing – Review \& Editing

\noindent\textbf{Duc-Tien Dang-Nguyen:} Writing – Review \& Editing

\noindent\textbf{Graham Healy:} Writing – Review \& Editing

\noindent\textbf{Steve Hodges:} Writing – Review \& Editing

\noindent\textbf{Björn Þór Jónsson:} Writing – Review \& Editing

\noindent\textbf{Luca Rossetto:} Conceptualization, Data Curation, Writing – Review \& Editing

\noindent\textbf{Klaus Schoeffmann:} Conceptualization, Data Curation, Writing – Review \& Editing

\noindent\textbf{Minh-Triet Tran:} Writing – Review \& Editing

\noindent\textbf{Lucia Vadicamo:} Conceptualization, Validation, Writing – Review \& Editing

\noindent\textbf{Cathal Gurrin:} Conceptualization, Funding acquisition, Writing – Review \& Editing

\bibliographystyle{plain}
\bibliography{ref,lsc24}

% ---------------------------------------------------------- %
\begin{IEEEbiography}[{\includegraphics[width=1in,height=1.25in,clip,keepaspectratio]{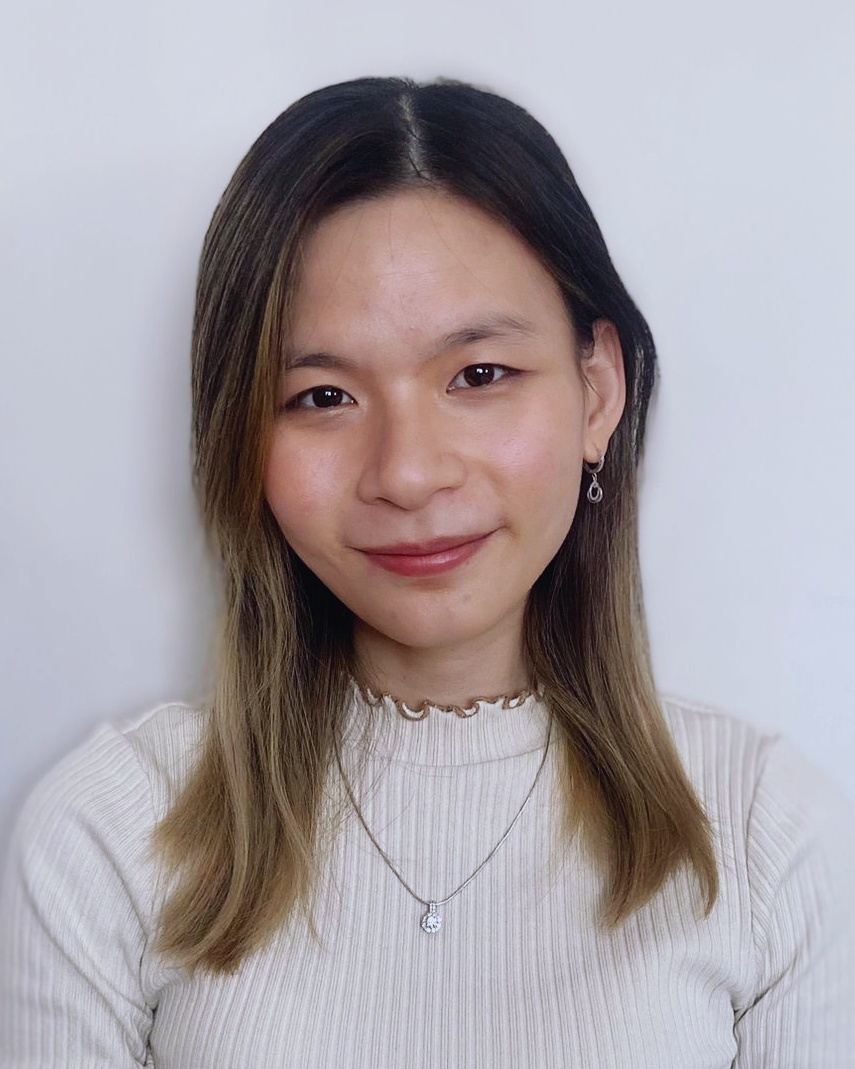}}]{Allie (Ly-Duyen) Tran} is a postdoctoral researcher at the School of Computing at Dublin City University, specialising in question answering over personal data with a focus on multimodal data understanding, exploration, and interaction. As an active member of the lifelog research community, she has contributed to several international benchmarking activities at LSC, NTCIR, and ImageCLEF\@. She is the principal developer of the MySc{\'e}al that has performed well at all recent instances of the LSC Challenge, and of the MyEachtra system specifically designed for QA tasks. She can be reached at \href{mailto:allie.tran@dcu.ie}{allie.tran@dcu.ie}.
\end{IEEEbiography}

% ---------------------------------------------------------- %

\begin{IEEEbiography}[{\includegraphics[width=1in,height=1.25in,clip,keepaspectratio]{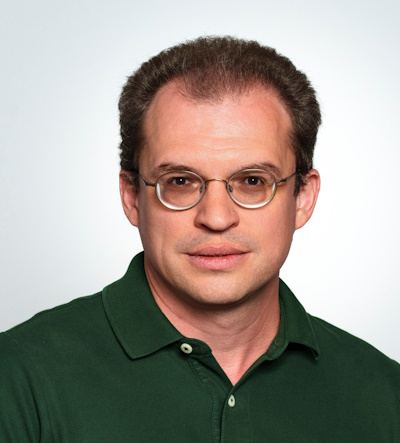}}]{Werner Bailer} is a Key Researcher at DIGITAL--Institute for Digital Technologies at JOANNEUM RESEARCH in Graz, Austria. He received a Dipl.-Ing.\@ degree in media technology and design in 2002, with a diploma thesis on motion estimation and segmentation. His research interests include audiovisual content analysis and retrieval, media production technologies and machine learning, contributing also to standardisation in these areas.
\end{IEEEbiography}

% ---------------------------------------------------------- %
\begin{IEEEbiography}
	[{\includegraphics[width=1in,height=1.25in,clip,keepaspectratio]{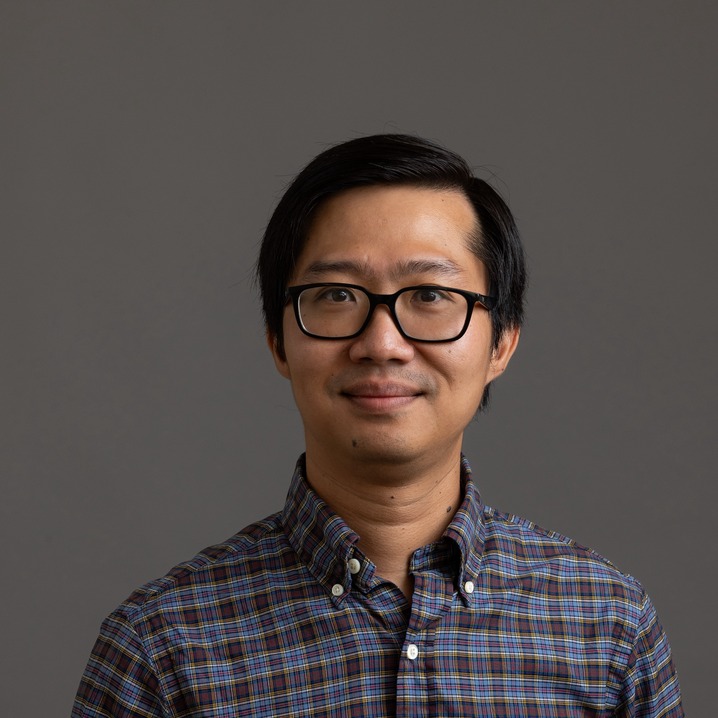}}] {Duc-Tien Dang-Nguyen} is a Professor at the Department of Information Science and Media Studies, University of Bergen, Norway. His research spans multimedia forensics, lifelogging, multimedia retrieval, and multimedia verification. He is the Partner-Lead (representing the University of Bergen) for The Nordic Observatory for Digital Media and Information Disorder (NORDIS), a European Horizon 2020-funded project (grant number 825469). Dang-Nguyen has authored over 160 peer-reviewed research papers in top-tier venues and has played a key role in organising more than 40 special sessions, workshops, and research challenges at ACM MM, ACM ICMR, NTCIR, ImageCLEF, and MediaEval over the past decade. His contributions to the academic community include serving as General Co-chair of MMM 2023 and CBMI 2025, TP Co-chair of MMM 2022, ACM ICMR 2024, and ACM MM 2025. He can be reached at \href{mailto:ductien.dangnguyen@uib.no}{ductien.dangnguyen@uib.no}.
\end{IEEEbiography}

% ---------------------------------------------------------- %
\begin{IEEEbiography}[{\includegraphics[width=1in,height=1.25in,clip,keepaspectratio]{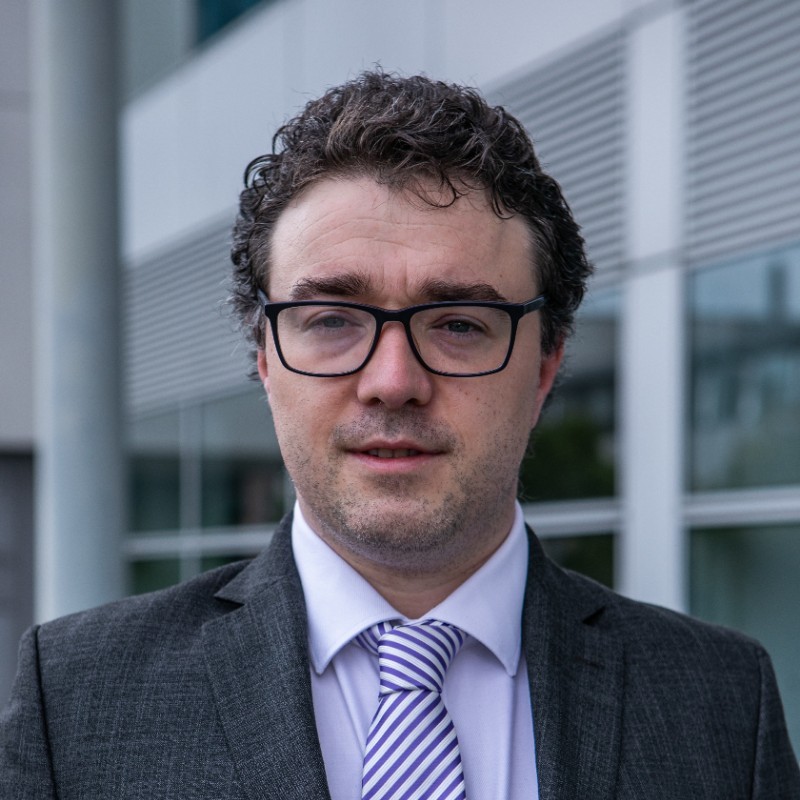}}]{Graham Healy} received the B.Sc\@. degree (Hons.) in computer applications, in 2008, and the PhD degree in brain–computer interfaces, in 2012. He is currently an Assistant Professor with the School of Computing, Dublin City University. His research interests lie in data analytics and human-computer interaction with a particular focus on novel interaction paradigms and data indexing/processing strategies. He can be reached at \href{mailto:graham.healy@dcu.ie}{graham.healy@dcu.ie}.
\end{IEEEbiography}

% ---------------------------------------------------------- %
\begin{IEEEbiography}
	[{\includegraphics[width=1in,height=1.25in,clip,keepaspectratio]{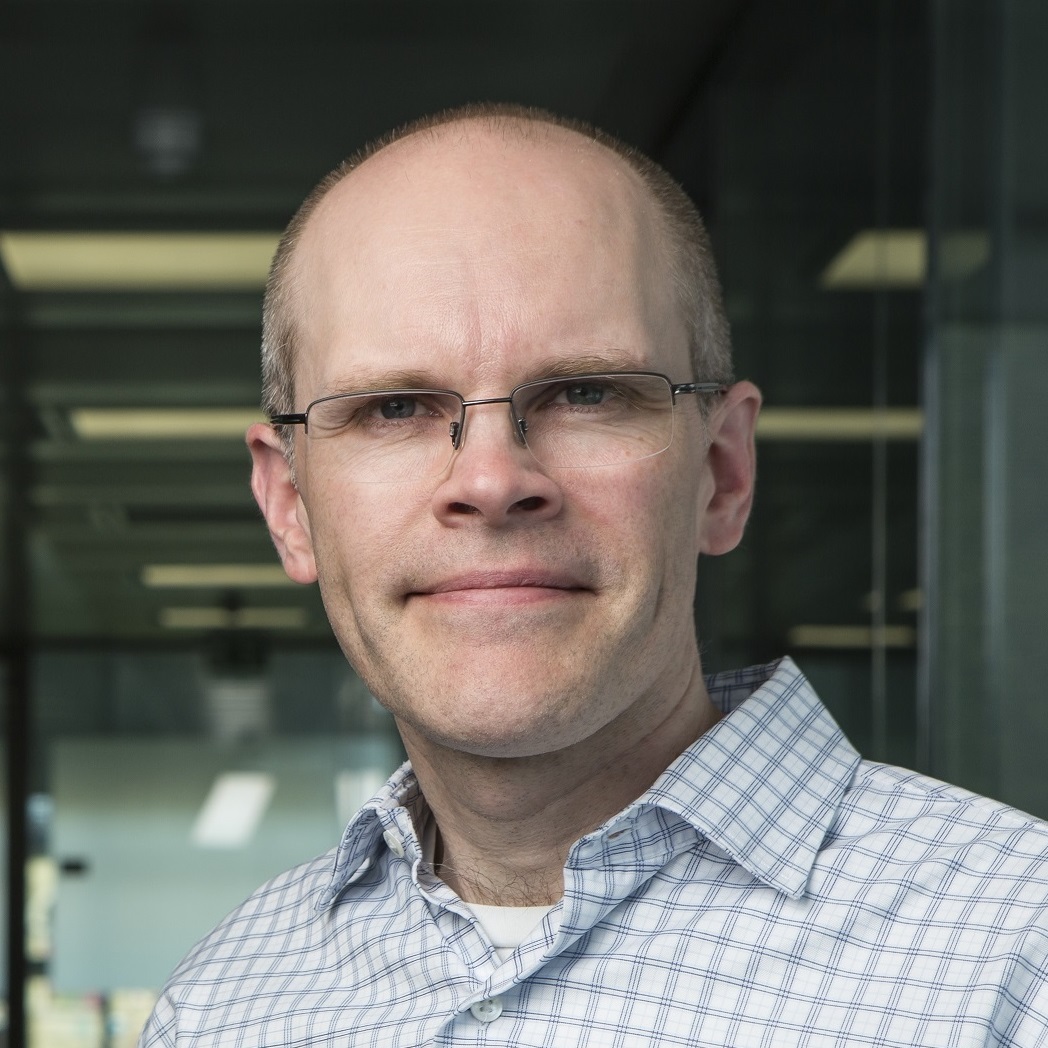}}] {Steve Hodges} is a Distinguished Professor at the School of Computing and Communications at Lancaster University, UK.\@ His research focuses on leveraging hardware innovations to make computers more engaging, inclusive, and useful. In the context of lifelogging, he is best known for his work on a pioneering wearable camera, the SenseCam. Steve is a Fellow of the IEEE and can be reached at \href{mailto:steve.hodges@lancaster.ac.uk}{steve.hodges@lancaster.ac.uk}.
\end{IEEEbiography}

% ---------------------------------------------------------- %
\begin{IEEEbiography}[{\includegraphics[width=1in,height=1.25in,clip,keepaspectratio]{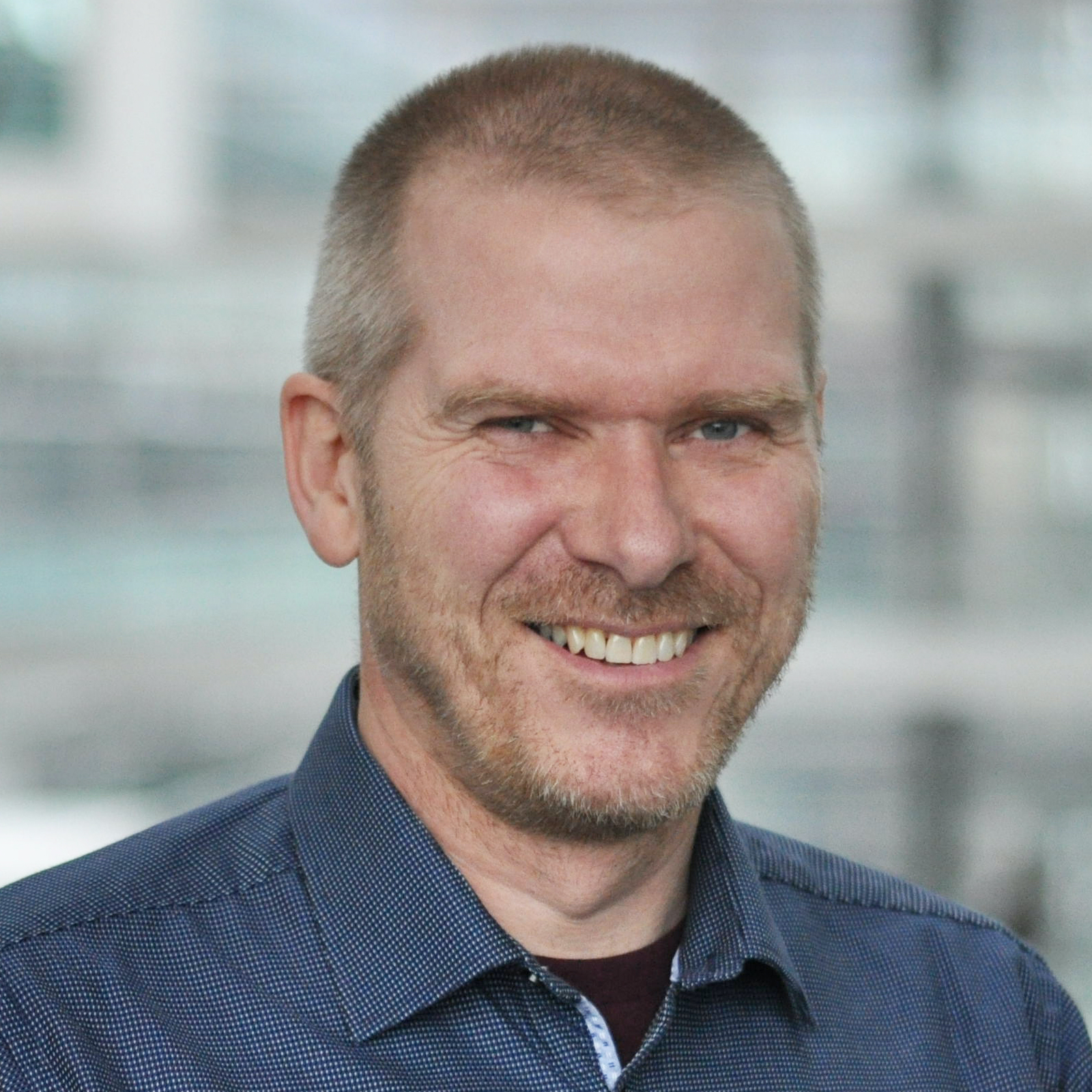}}]{Bj{\"o}rn {\th}{\'o}r J{\'o}nsson} is a Full Professor in the Department of Computer Science at Reykjavik University, Iceland. He received a B.Sc.\@ degree in Computer Science from the University of Iceland in 1991, and a PhD degree in Computer Science from the University of Maryland, College Park, USA, in 1999.  His research work focuses primarily on the performance and scalability of content-based multimedia analytics and retrieval.
\end{IEEEbiography}

% ---------------------------------------------------------- %
\begin{IEEEbiography}[{\includegraphics[width=1in,height=1.25in,clip,keepaspectratio]{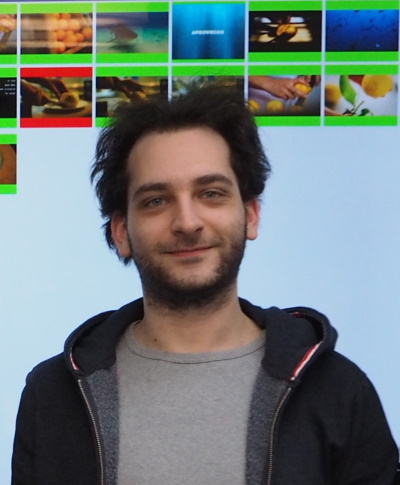}}]{Luca Rossetto} is currently an Assistant Professor at the School of Computing at Dublin City University. His research focuses on managing, analysing, and retrieving multimodal data. He is one of the core developers of the open-source multimedia retrieval engine ‘vitrivr’ and co-creator of the ‘Distributed Retrieval Evaluation Server’ used for interactive multimedia evaluations in different areas. Luca is a member of ACM and SIGMM and a regular reviewer for international multimedia conferences such as ACM MM, ACM MM Asia, ACM ICMR, MMM, and others, as well as journals including IEEE Transactions on Multimedia, Multimedia Systems, and Multimedia Tools and Applications.
\end{IEEEbiography}

% ---------------------------------------------------------- %
\begin{IEEEbiography}[{\includegraphics[width=1in,height=1.25in,clip,keepaspectratio]{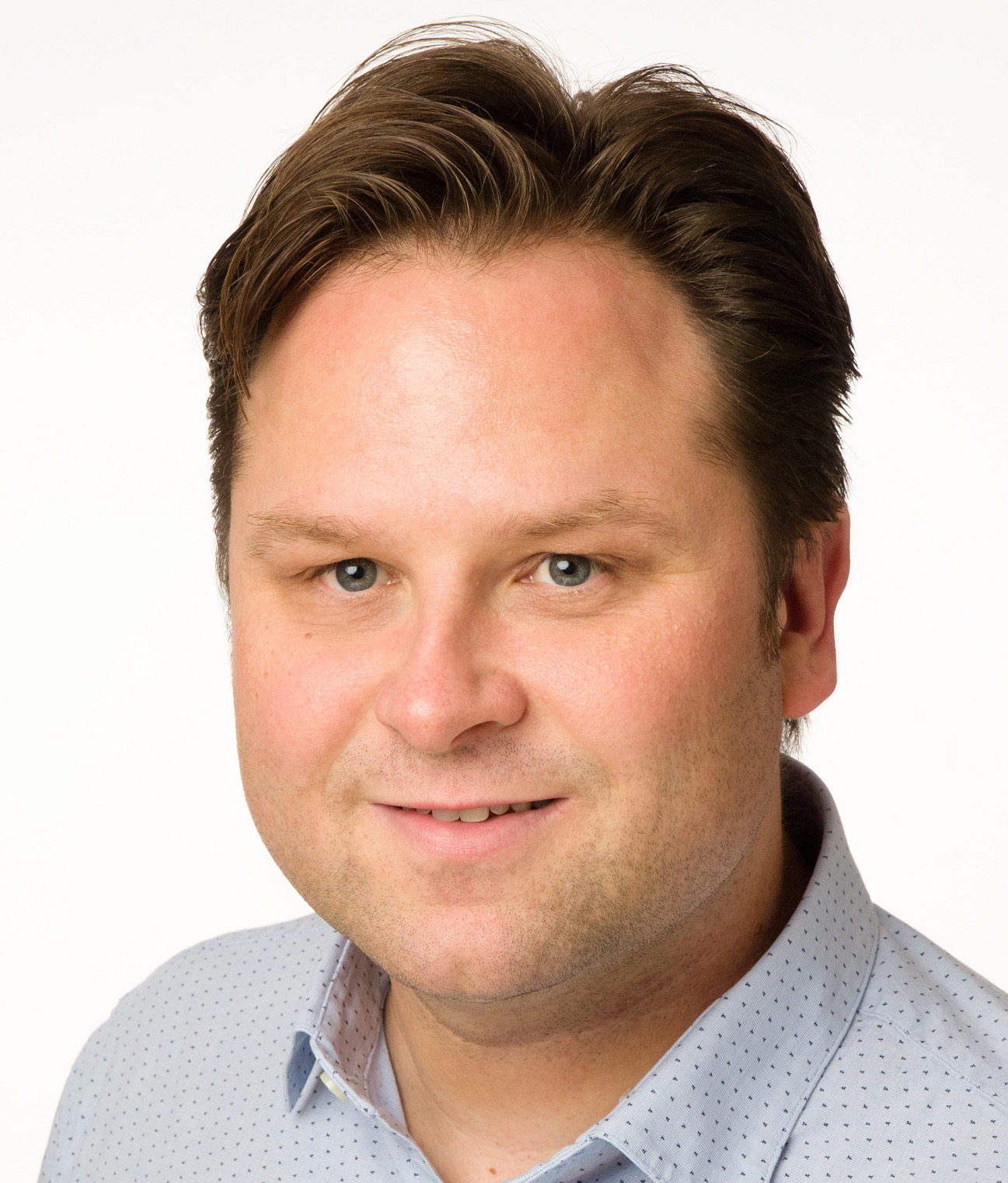}}]{Klaus Schoeffmann} is an Associate Professor at the Institute of Information Technology (ITEC) at Klagenfurt University, Austria. He holds a PhD and an MSc in Computer Science. His research focuses on video content understanding (particularly medical/surgery videos), multimedia retrieval, interactive multimedia, and applied deep learning. He is one of the developers of LIFEXPLORE and can be reached at \href{mailto:Klaus.Schoeffmann@aau.at}{Klaus.Schoeffmann@aau.at}
\end{IEEEbiography}

% ---------------------------------------------------------- %
\begin{IEEEbiography}[{\includegraphics[width=1in,height=1.25in,clip,keepaspectratio]{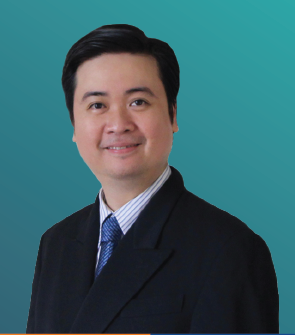}}]{Minh-Triet Tran} received the B.Sc., M.Sc., and PhD degrees in computer science from the University of Science, VNU-HCM, in 2001, 2005, and 2009. In 2001, he joined the University of Science. He was a Visiting Scholar with the National Institutes of Informatics (NII), Japan, from 2008 to 2010, and the University of Illinois at Urbana–Champaign (UIUC), from 2015 to 2016. His research interests include cryptography, security, computer vision, and human–computer interaction.
	He is currently the Vice President of the University of Science, VNU-HCM\@. He is also the Membership Development, Student Activities Coordinator of IEEE Vietnam. He is also a member of the Advisory Council for Artificial Intelligence development of Ho Chi Minh City, and Vice Chairman of Vietnam Information Security Association (VNISA, South Branch).
	He can be reached at \href{mailto:tmtriet@hcmus.edu.vn}{tmtriet@hcmus.edu.vn}.
\end{IEEEbiography}

% ---------------------------------------------------------- %
\begin{IEEEbiography}[{\includegraphics[width=1in,height=1.25in,clip,keepaspectratio]{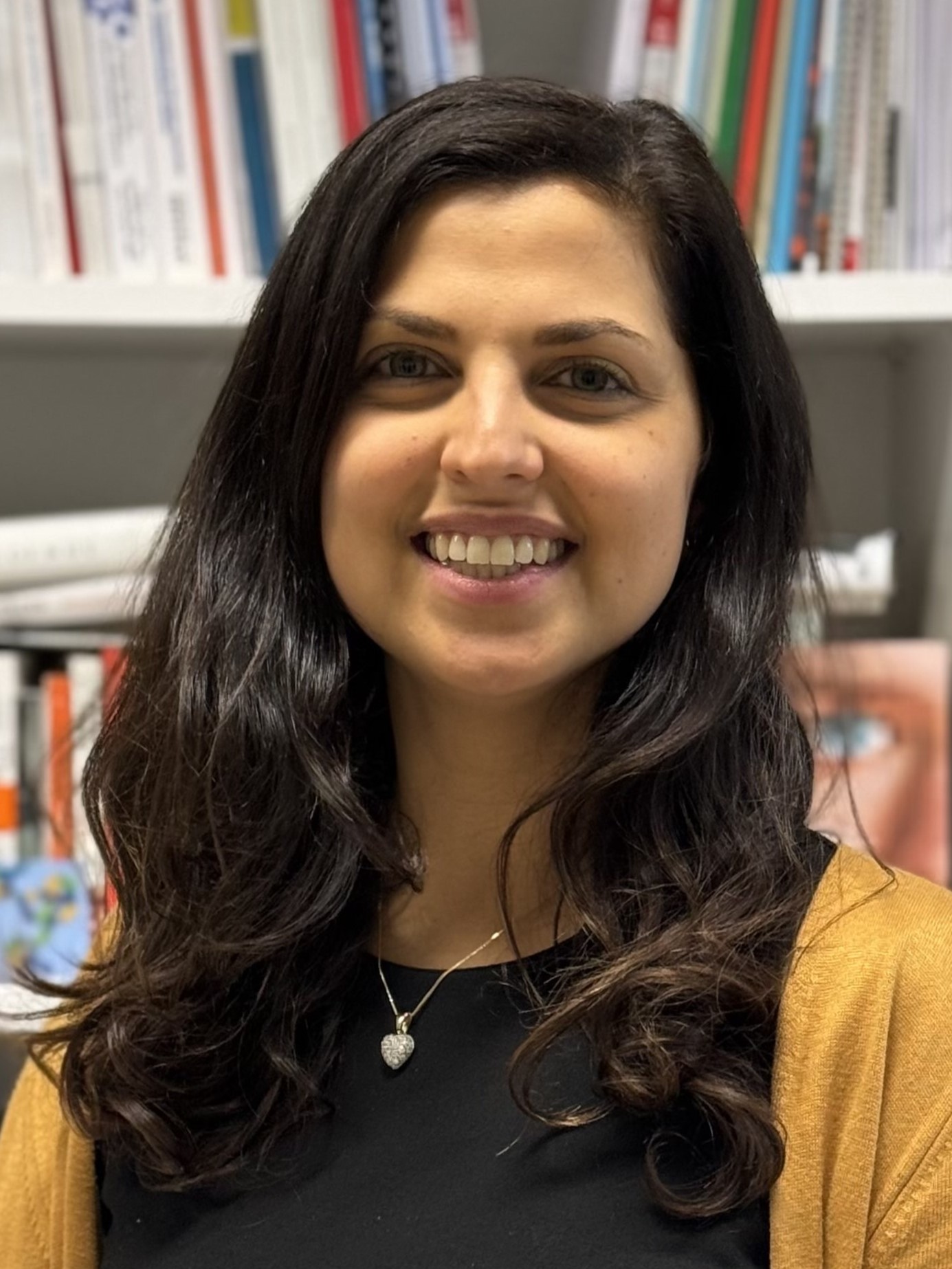}}]{Lucia Vadicamo} is a senior researcher at the Information Science and Technologies Institute (ISTI) of the National Research Council (CNR) in Pisa, Italy. She holds a PhD in Information Engineering (2018) and a Master's degree in Mathematics (2013) from the University of Pisa, both with honours. Her research activity focuses on Multimedia Information Retrieval, Artificial Intelligence, and Similarity Search. She is leading a team dedicated to the development of the VISIONE system.
\end{IEEEbiography}

% ---------------------------------------------------------- %
\begin{IEEEbiography}[{\includegraphics[width=1in,height=1.25in,clip,keepaspectratio]{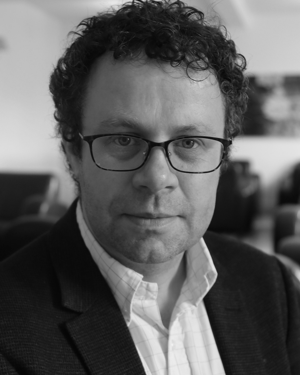}}]{Cathal Gurrin} received a BSc in Computer Applications from Dublin City University in 1997 and a PhD from Dublin City University in 2002.

	He is a co-founder of the LSC Lifelog Search Challenge workshop and the NTCIR-Lifelog collaborative benchmarking activity. He was the co-author of Lifelogging: Personal Big Data, and his research has been covered internationally on the BBC, Discovery Channel, as well as in print media (Economist magazine, New York Times).
\end{IEEEbiography}

% ---------------------------------------------------------- %
\EOD
\end{document}